\documentclass[twocolumn,showpacs,amsmath,pra,aps,amssymb,superscriptaddress,nofootinbib]{revtex4-1} 
\usepackage[T1]{fontenc}
\usepackage[latin9]{inputenc}
\usepackage{times}
\usepackage{color} 
\usepackage{xspace}
\usepackage{amssymb,amsmath}
\usepackage{amsbsy}
\usepackage[pdftex]{graphicx}
\usepackage{bm}
\usepackage{float}
\usepackage{adjustbox}
\usepackage{epstopdf}

\usepackage[unicode,breaklinks]{hyperref}
\hypersetup{ 
    unicode=true,
    plainpages=false, 
    colorlinks=true,
    linkcolor=blue,
    citecolor=blue,
    filecolor=black,
    urlcolor=blue
}
\usepackage{url}
\usepackage{verbatim}

\begin{document}

\title{Different growth rates for spin and superfluid order in a quenched spinor condensate}
\author{Andr{\'e}ane Bourges}  
\affiliation{{\'E}cole Normale Sup{\'e}rieure de Lyon, Universit{\'e} Claude Bernard Lyon I, France}
\affiliation{Quantum Science Otago and the Dodd-Walls Centre for Photonic and Quantum Technologies, Department of Physics, University of Otago, Dunedin 9016, New Zealand}
\author{P.~B.~Blakie}  
\affiliation{Quantum Science Otago and the Dodd-Walls Centre for Photonic and Quantum Technologies, Department of Physics, University of Otago, Dunedin 9016, New Zealand}
\date{\today}
\begin{abstract}
 In this paper we study the coarsening dynamics of a spinor condensate quenched into an easy-axis ferromagnetic phase by a sudden change in the quadratic Zeeman energy. We show that applying a spin rotation prior to changing the Zeeman energy accelerates the development of local order and reduces heating. We examine the longitudinal spin ordering and the superfluid ordering of the system and show that the respective order parameter correlation functions exhibit dynamic scaling in the late time dynamics. Our results also demonstrate that these two types of order grow at different rates, i.e.~with different dynamic critical exponents. The  spin domain area distribution is calculated and is shown to have power law scaling behavior expected from percolation theory.  
\end{abstract}

\maketitle

\section{Introduction}
A spinor Bose-Einstein condensate is a system that possesses both superfluid and magnetic order \cite{Kawaguchi2012R,StamperKurn2013a}. Experiments are able to explore transitions between phases of different magnetic order by controlling the quadratic Zeeman energy ($q$) that shifts the magnetic sublevels \cite{Sadler2006a,Leslie2009a,Bookjans2011b,Guzman2011a,De2014a}. Such a system presents a rich playground for studies of non-equilibrium phenomena, including defect formation related to the rate that the phase transition is crossed \cite{Sadler2006a,Lamacraft2007a,Saito2007a,Saito2007b,Uhlmann2007a,Damski2007a,Bookjans2011b,Vinit2013a,Witkowska2014a}, through to the late-time coarsening dynamics describing how the small domains produced by the quench anneal towards the equilibrium state \cite{Mukerjee2007a,Lamacraft2007a,Vengalattore2010a,Guzman2011a,Kudo2013a,Kudo2015a,Williamson2016a,Williamson2016b}. Studies have largely focused on the spin properties of the system, leaving the superfluid dynamics unexplored. We note that in binary condensates, which can be regarded as a pseudo spin-$\frac{1}{2}$ system, there has been work on both spin coarsening \cite{Hofmann2014} and aspects of the superfluid behavior (e.g.~analyzed via kinetic energy spectra  \cite{Karl2013a}).
The development of various techniques for measuring  spin \cite{Higbie2005a,Vengalattore2010a,Guzman2011a} and superfluid correlations (e.g.~see \cite{Bloch2000a,Donner2007a,Clade2009a,Navon2015a,Chomaz2015a}) demonstrate that it is possible to study the evolution of both types of order in spinor experiments.  The relationship between spin and superfluid order has been considered in the equilibrium properties of spinor condensates. For example, in Ref.~\cite{Natu2010a} it was found that for a large spin-dependent interaction, ferromagnetism emerges at a higher temperature than condensation. 
 
Here we study the superfluid and spin ordering dynamics of a ferromagnetic spin-1 condensate quenched into a easy-axis ferromagnetic phase, where the spin order preferentially aligns (or anti-aligns) along the quantization axis defining the quadratic Zeeman energy shift.  Initially small domains form, but grow as time passes.
 At sufficiently late times this coarsening dynamics can enter a universal scaling regime  \cite{Bray1994}:  the correlation function of the order parameter collapses to a universal (time-independent) form when space is scaled by a characteristic length $L(t)\sim t^{1/z}$, which yields the dynamic critical exponent $z$. Here we find that the spin and superfluid order exhibit universal scaling, but do not develop identically. The spin order spreads more rapidly across the system than the superfluid order, with different dynamic critical exponents for each type of ordering.  
 
 In this paper we consider two types of quenches. The first (standard) quench (e.g.~see \cite{Sadler2006a,Kudo2013a,Williamson2016a}) is to start with an equilibrium non-magnetic (polar) condensate at large $q$ and then to suddenly quench to a negative $q$ value where the easy-axis magnetic phase is the ground state. We also propose a second type of quench that utilizes the atomic physics toolbox of coherent manipulations: this differs from the first quench in that a $\pi/2$ spin rotation is applied to the initial state immediately prior to $q$ being quenched. 
Both quenches produce the same late time coarsening behavior, but we show that the second quench has less heating and the early-time dynamics (where local order develops) concludes more rapidly, suggesting this may be more suitable for experiments where long times are difficult to access. 
We finally consider the areas of the individual spin domains that form after the quench and how these evolve in time. Our results show that once the domains develop in the system their areas vary over many orders of magnitude. At sufficiently long times the domain area distribution decays as a power law $\sim S^{-2}$ for sufficiently large $S$, where $S$ is the domain area. This can be related to ideas of percolation theory \cite{Stauffer1994}, as has also been pointed in in recent work on the immiscibility phase transition in a binary condensate \cite{Takeuchi2015a,Takeuchi2016a}. 

The outline of the paper is as follows. In Sec.~\ref{SecFormalism} we introduce the dynamical formalism used to simulate the quench dynamics. We introduce the two different quench types considered and  discuss the role and nature of the unstable excitations that drive the initial phase transition dynamics. Our main results are presented in Sec.~\ref{SecResults}. First we consider the early-time dynamics to emphasize the differences in how local spin order emerges for the two quench types. We then turn to considering the spin and superfluid order, as described by the relevant two-point correlation functions. We demonstrate that in the late time dynamics these correlation functions exhibit dynamical scaling, albeit with different growth laws. Then we compute the spin domain areas, extract the domain size distribution and investigate the scaling of this distribution as a function of domain size and time.  Finally we conclude in Sec.~\ref{SecConclude}.

\section{Formalism}\label{SecFormalism}
\subsection{The Spin-1 Gross-Pitaevskii equations} 
The system we consider is a homogeneous quasi-two-dimensional (quasi-2D) spin-1 condensate  
described by the Hamiltonian~\cite{Ho1998a,Ohmi1998a}
\begin{align}\label{spinH}
H\!=\!\int\!d^2\bm{x}\left[\bm{\psi}^\dagger\!\left(\!-\frac{\hbar^2\nabla^2}{2M}+qf_z^2\right)\!\bm{\psi}+\frac{g_n}{2}n^2+\frac{g_s}{2}\left|\bm{F}\right|^2\right]\!.
\end{align}
Here $\bm{\psi}\equiv (\psi_{1},\psi_0,\psi_{-1})^T$ is a three component spinor describing the condensate amplitude in the three spin hyperfine sublevels  ($m=+1,0,-1$) and $q$ is the quadratic Zeeman shift\footnote{This linear Zeeman shift can be removed by transforming to a rotating frame and we neglect this here.} arising from the presence of an external field along $z$.  The interactions are described by a density dependent term $g_nn^2$  and a spin-density dependent term $g_s|\bm{F}|^2$, where $g_n$ and $g_s$ are density-dependent  and spin-dependent coupling constants, $n\equiv\bm{\psi}^\dagger\bm{\psi}$ is the total number density, and $\bm{F}\equiv \bm{\psi}^\dagger\bm{f}\bm{\psi}$  is the spin density, with $(f_x,f_y,f_z)\equiv\bm{f}$ being the spin-1 matrices.  
The dynamics of the system is  described by the spin-1 Gross-Pitaevskii equation (GPE)
\begin{align}\label{spinGPEs}
i\hbar\frac{\partial\bm{\psi}}{\partial t}=\left(-\frac{\hbar^2\nabla^2}{2M}+qf_z^2+g_nn+g_s\bm{F}\cdot\bm{f}\right)\bm{\psi}.
\end{align}

For the system to be mechanically stable we require $g_n>0$, and we additionally restrict our attention here to the case of ferromagnetic interactions, i.e.~$g_s<0$, as realized in $^{87}$Rb  condensates \cite{Chang2004a}.  In spinor condensate experiments the quasi-2D regime has been realized by using a trapping potential with tight confinement in one direction (e.g.~see \cite{Sadler2006a}). Our interest is in homogeneous systems where the phase transition dynamics are simpler, noting that recent experiments have realized flat-bottomed optical traps for this purpose \cite{Navon2015a,Chomaz2015a} (also see~\cite{damle1996b}).

\subsection{Ground state phases and quenches}\label{SecPhasesQuenches}

\begin{figure}
\centering
\includegraphics[width=0.5\textwidth]{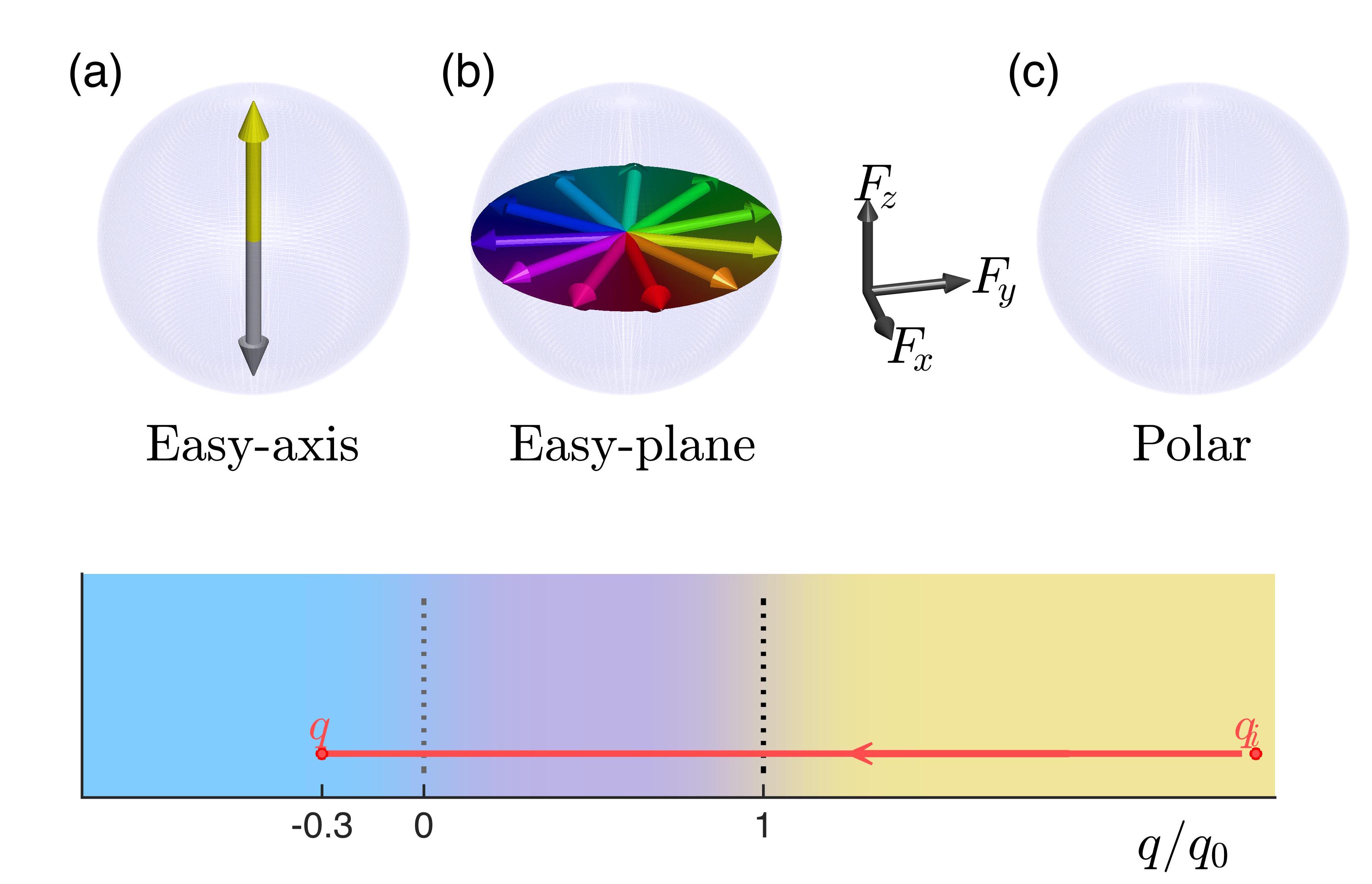}
\caption{\label{phaseDiag}  Magnetic phase diagram for a ferromagnetic spin-1 condensate with zero $z$-magnetization. The spheres show the direction of magnetization in the three states. (a) For $q<0$ the magnetization lies along the $F_z$ axis and the state is termed \emph{easy-axis}. (b) For $0<q<q_0$ the magnetization lies in the transverse ($F_x$-$F_y$) plane and the state is termed \emph{easy-plane}. Quenches to this phase are not analyzed here but can be found in Refs.~\cite{Barnett2011,Kudo2015a,Williamson2016a}. (c) For $q>q_0$ the $m=\pm 1$ levels are unoccupied and the system is unmagnetized. This state is termed \emph{polar}. The directed line on the phase diagram indicates the quench we consider in this paper, where the Zeeman energy is suddenly reduced from an initial value $q_i>q_0$ where the ground state is polar, to a value $q<0$ where the system favours easy-axis spin ordering.}
\end{figure}

The ground state of Eq.~(\ref{spinH}) depends on the value of $q$ (e.g.~see \cite{Kawaguchi2012R}) relative to the characteristic spin energy $q_0=2|g_s|n_0$, where  $n_0$ is the (uniform) condensate density. A schematic phase diagram and representation of the ground states important to this paper are shown in Fig.~\ref{phaseDiag}. There are two ferromagnetic phases (a) and (b), which differ in their symmetries, and a non-magnetized polar phase (c). Of the two ferromagnetic phases we only consider the easy-axis phase here as this phase admits well-defined spin domains.  We briefly review the relevant order parameters for our simulations, which we express in the form  
\begin{equation}
\bm{\psi}=e^{i\theta}\sqrt{n_0}\bm{\xi},
\end{equation}
where   $\theta$ is a global (superfluid) phase, and $\bm{\xi}=(\xi_{1},\xi_0,\xi_{-1})^{T}$ is a normalized spinor (i.e.~$\bm{\xi}^\dagger\bm{\xi}=1$).  
For high values of the quadratic Zeeman energy ($q>q_0$) the the ground state is the non-magnetized polar phase, with normalized spinor  
\begin{align}
\bm{\xi}_{\mathrm{P}}=\left(\begin{array}{c}0\\1\\0\end{array}\right),
 \end{align}
for which $\mathbf{F}=\mathbf{0}$. This state, or a spin rotated form of this state, is the initial condition for our quench.
Our interest is how the system then reorders when $q$ is quenched to a negative value, where the ground state is an easy-axis ferromagnetic phase with normalized spinor
\begin{align}
\bm{\xi}_{\mathrm{EA}}=\left(\begin{array}{c}1\\0\\0\end{array}\right)\quad\mbox{or}\quad\left(\begin{array}{c}0\\0\\1\end{array}\right),
\end{align}
corresponding to magnetization of +1 or -1 along $z$, respectively. These states break the $\mathbb{Z}_2$ ($z$-reflection) symmetry of the Hamiltonian. 
  
We now introduce the two types of quench we use to transition the system from the polar to the easy-axis phase.\\
\textbf{Quench 1 (Q1)}: In this quench the quadratic Zeeman energy is simply set to a negative value at $t=0$. Here the  polar state is dynamically unstable and decays, initially by developing transverse magnetization. In this case an energy per particle of
\begin{equation}
\Delta\epsilon=(\tfrac{1}{4}
q_0-q),\end{equation}
is liberated (i.e.~the excess energy that the polar state has over the easy-axis ferromagnetic state for $q<0$), and available to heat the system.\\
\textbf{Quench 2 (Q2)}: In this quench a spin rotation of $U_{\mathrm{rot}}=e^{-i\frac{\pi}{2} f_x}$, 
is performed producing a so-called anti-ferromagnetic state:
\begin{equation}
\bm{\xi}_{\mathrm{AF}}=U_{\mathrm{rot}}\bm{\xi}_{\mathrm{P}}=\frac{1}{\sqrt{2}}\left(\begin{array}{c}-1\\0\\1\end{array}\right). 
\end{equation}
After the spin rotation has been preformed the  quadratic Zeeman energy is  immediately set to a negative value. We note that this rotation (using RF pulses) and quench to negative $q$ in a  spin-1 experiment has been reported in Ref.~\cite{Seo2015a}.
 
For this initial condition the energy per particle available to heat the system is smaller than for Q1 and is independent of $q$ (for $q<0$), i.e.~
\begin{equation}
\Delta\epsilon=\tfrac{1}{4}q_0.
\end{equation}

\subsection{Excitations of unstable initial states}\label{Secunstablemodes}
\begin{figure}
\centering
\includegraphics[width=0.5\textwidth]{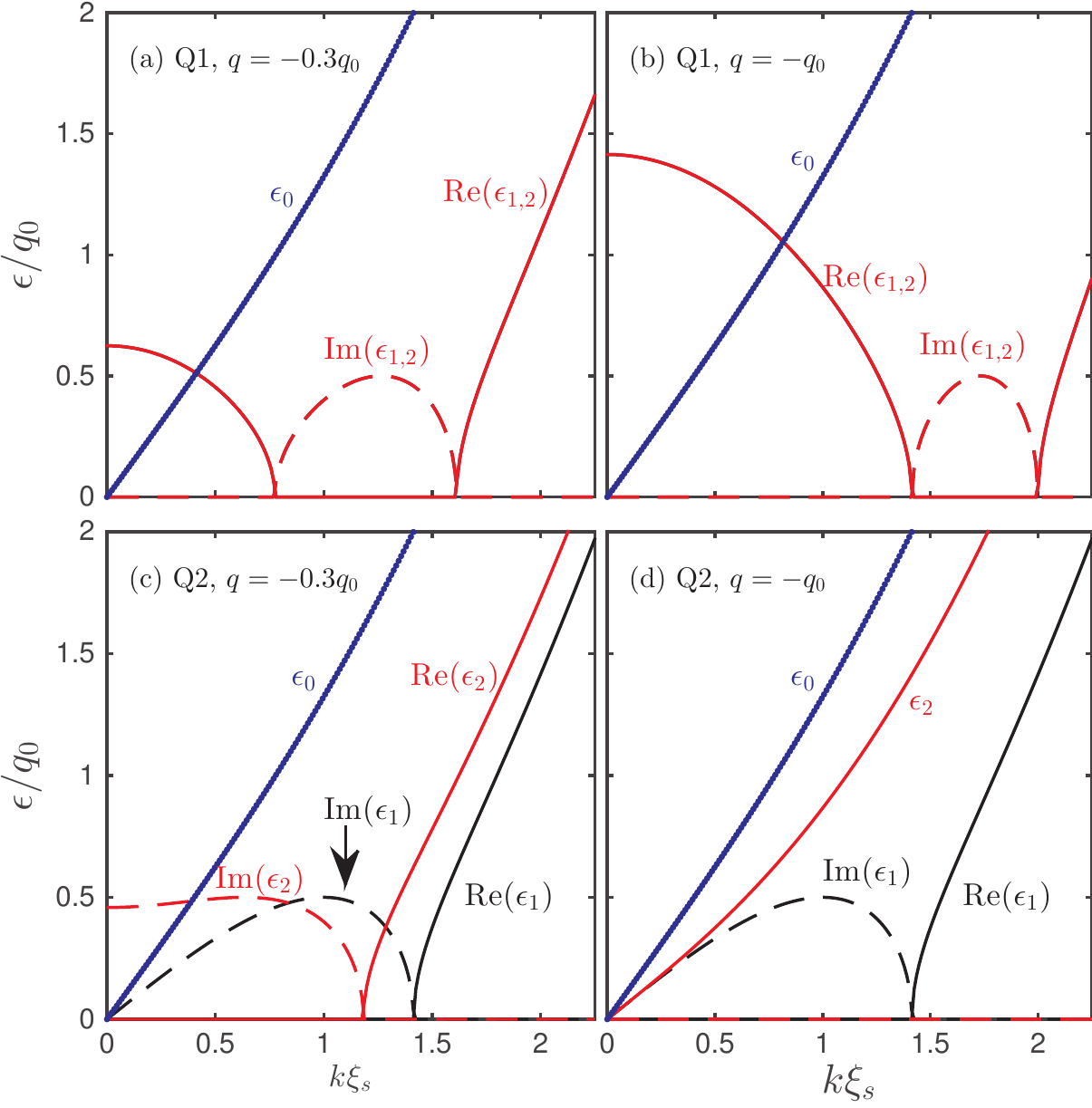}
\caption{\label{Figspectrum} (Color online) Bogoliubov spectra for the post-quench initial states emphasizing the dynamically unstable modes. Spectra of the $\bm{\xi}_{\mathrm{P}}$ initial state produced by the Q1 quench at (a) $q=-0.3q_0$ and (b) $q=-q_0$.  Spectra of the $\bm{\xi}_{\mathrm{AF}}$ initial state produced by the Q2 quench at (a) $q=-0.3q_0$ and (b) $q=-q_0$.  The lines indicate the real parts of the energies, while the dashed lines indicate the imaginary part. The phonon  ($\epsilon_0$) and magnon ($\epsilon_1,\epsilon_2$) excitation branches are labelled in the plots, noting that the magnon branches are degenerate in (a) and (b). These results are for the case $g_s=-g_n/3$. }
\end{figure}

Insight into the  initial post-quench dynamics is also provided through understanding the Bogoliubov quasi-particle excitations of the initial state. For a spin-1 system there are three excitation branches, which can be typically classified as a phonon branch and two magnon (or spin-wave) branches. In discussing these modes we adopt the naming conventions used in \cite{Symes2014a} and refer to the phonon branch as $\nu=0$ and the magnon branches as $\nu=1,2$. The instabilities of the initial state following the quench are revealed by the imaginary parts in the quasi-particle energies $\epsilon_\nu(k)$. As our system is mechanically stable (and $g_n\gg|g_s|$) the phonon branch always remains stable, and exhibits a rapidly rising linear spectrum $\epsilon_0\sim c_n\hbar k$, where $c_n=\sqrt{g_nn_0/M}$ is the speed of sound.

\subsubsection{Unstable magnons for the Q1 quench}\label{SecmagnonsQ1}
For the Q1 quench the initial non-equilibrium state  (P-phase) at $q<0$ has a Bogoliubov spectrum with two unstable magnon branches. For our case of zero magnetization along $z$ these magnon branches\footnote{The spin-1  Bogoliubov spectra are reviewed in Ref.~\cite{Kawaguchi2012R}, and are given here specific for the case $p=0$ and a condensate with zero $z$-magnetization. } are degenerate with the dispersion relation 
\begin{equation}
\epsilon_{1,2}(k)=\sqrt{(\epsilon_k^0+q)(\epsilon_k^0+q+2g_sn_0)},\label{EqMagnonP}
\end{equation}
 where $\epsilon_k^0=\hbar^2k^2/2M$. For $q<0$ these magnon branches have dynamically unstable modes (i.e.~the quasi-particle energies $\epsilon_{1,2}$ are imaginary) for the range of $k$ values
 \begin{equation}
 \sqrt{-2q/q_0}<  k\xi_s<\sqrt{2(1-q/q_0)},\qquad(q<0).
 \end{equation}
These magnon modes have amplitude in the $m=\pm1$ sub-levels and lead to the decay of the polar phase by the development of transverse magnetization. As the value of $q$ changes the character of the unstable modes remains qualitatively the same [see Figs.~\ref{Figspectrum}(a) and (b)], with the quantitative change that the unstable modes shift to larger $k$ as $q$ becomes more negative.

\subsubsection{Unstable magnons for the Q2 quench}\label{SecmagnonsQ2}
For the Q2 quench the initial  state  (AF-phase) has two magnon branches of qualitatively different behavior 
\begin{align}
\epsilon_{1}(k)&=\sqrt{\epsilon_k^0(\epsilon_k^0+2g_sn_0)},\label{EqMagnonAF1}\\
\epsilon_{2}(k)&=\sqrt{(\epsilon_k^0-q+g_sn_0)^2-(g_sn_0)^2}.\label{EqMagnonAF2}
\end{align}
The $\nu=1$ branch consists of modes that have amplitude in the $m=\pm1$ sublevels giving rise to longitudinal magnetic fluctuations, while the $\nu=2$ branch modes have amplitude in the $m=0$ sublevel giving rise to transverse fluctuations.   The unstable modes for the $\nu=1$ branch are independent of $q$ and occur for the $k$ range
\begin{equation}
k<\sqrt{2}/\xi_s,
\end{equation}
e.g., see Figs.~\ref{Figspectrum}(c) and (d).
The $\nu=2$ branch has a finite region of unstable modes 
 \begin{equation}
\mathrm{Re}\{\sqrt{2q/q_0}\}<  k\xi_s<\mathrm{Re}\{\sqrt{2(1+q/q_0)}\},
 \end{equation}
for $q>-q_0$ (noting that the lower bound is 0 for $q<0$).  At $q\le-q_0$ [see Figs.~\ref{Figspectrum}(d)], this branch becomes stable, and the condensate dynamics is driven entirely by the unstable modes in the $\nu=1$ branch.

\subsection{Details of simulation method}\label{SecSimDetails}

We simulate the dynamics using the spin-1 GPE (\ref{spinGPEs}) with noise added to the initial state to seed the growth of symmetry breaking domains. 
For the initial state of the Q1 quench we take
\begin{align}
\bm{\psi}(\bm{x})=\sqrt{n_0}\bm{\xi}_{\mathrm{P}}+\bm{\delta}(\bm{x}),\label{EqPIC}
\end{align}
where $\sqrt{n_0}\bm{\xi}_{\mathrm{P}}$ is a uniform (zero-momentum) polar condensate and $\bm{\delta}$ is a small noise field used to seed the dynamical instabilities that occur after the quench. The precise form of noise used is based on the truncated Wigner formalism \cite{cfieldRev2008} and is described in Ref.~\cite{Williamson2016b}. For the Q2 quench the spin rotation $U_{\mathrm{rot}}$ is applied to state (\ref{EqPIC}) to yield the initial condition.

We use a condensate areal density of  $n_0=10^4/\xi_s^2$, where $\xi_s\equiv\hbar/\sqrt{q_0 M}$ is the spin healing length. We evolve the spin-1 GPE for the spinor field $\bm{\psi}$ on a 2D square grid with dimensions $l\times l$ covered by an $N\times N$ grid of equally spaced points. Here most of our results are calculated on a  grid of linear size $l=510\xi_s$ with $N=1024$ points. 
We evolve the spin-1 GPE~\eqref{spinGPEs} using a fourth-order symplectic method that uses Fast Fourier transforms to evaluate the kinetic energy operators with spectral accuracy \cite{Symes2016a}.
  The quadratic Zeeman energy is set at the final quench value $q<q_0$ for the duration of the simulation dynamics, so that the quench is effectively instantaneous at $t=0$ from the initial condition for the case of the Q1 and Q2 quench, respectively. In order for the post-quench dynamics to become universal it is necessary to simulate the system over many spin times $t_s\equiv \hbar/q_0$. Indeed, here we evolve out to times of up to $t=10^4\,t_s$, and on such long time-scales the use of a highly accurate symplectic algorithm ensures we obtain accurate solutions that conserve energy and magnetization. 
  All the results presented here are for $g_s=-\tfrac{1}{3}g_n$. This choice is made to ensure our interactions are in a similar regime to the binary condensate immiscibility simulations of Ref.~\cite{Takeuchi2016a}, noting that the mapping of binary condensate parameters onto effective density and spin dependent interactions is discussed in Ref.~\cite{StamperKurn2013a}. We have also performed simulations for $g_s=-\frac{1}{12}g_n$ to verify that our results are qualitatively  unchanged with a smaller spin-dependent interaction. 
 
\section{Results}\label{SecResults}

\subsection{Early-time dynamics}\label{SecEarlyTime}
 \begin{figure}
\centering
\includegraphics[width=0.5\textwidth]{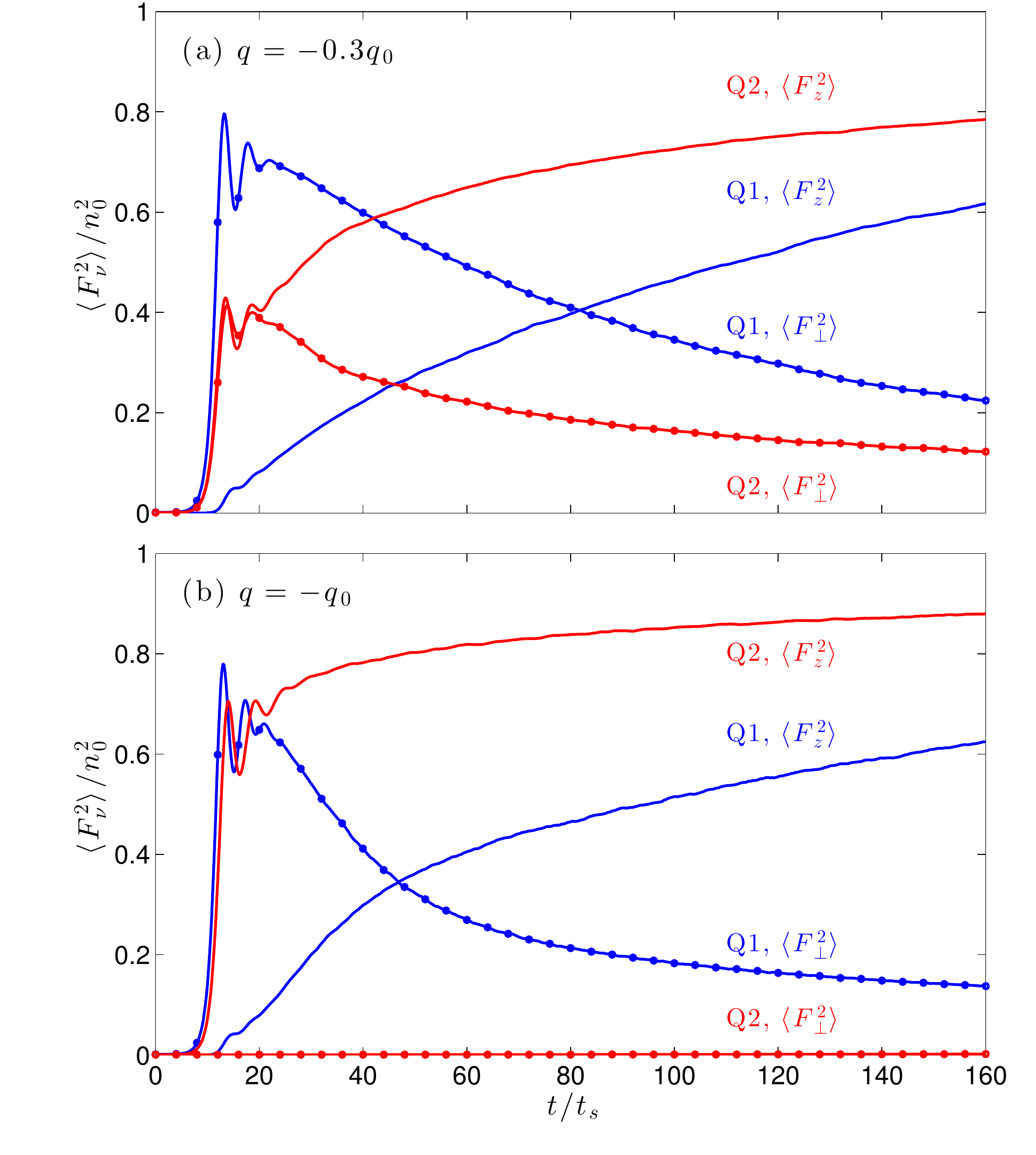}
\caption{\label{Figstdynamics}   Growth of local magnetization following the  Q1 and  Q2  quench to (a) $q=-0.3q_0$ and (b) $q=-q_0$.  In all cases the (local) magnetization is calculated as a spatial average over the system at each time, i.e.~$\langle F_\nu^2\rangle= l^{-2}\int d^2\mathbf{x}\,F^2_\nu(\mathbf{x}) $ for $\nu=z$ (plain lines) and $\nu=\perp$ (lines with symbols).  Other parameters as in Fig.~\ref{Figspectrum}. }
\end{figure}
For both the Q1 and Q2 quenches the initial state is unmagnetized, but the unstable modes lead to the spin density growing exponentially. The initial post-quench dynamics depends upon the kind and depth (i.e.~$q$)  of the quench.  We take the \textit{early-time} regime to be the initial time period over which the longitudinal magnetization locally develops. For our parameters (see Fig.~\ref{Figstdynamics}) this regime extends up to $t\sim10^2\,t_s$. For times after this the longitudinal  magnetization is the dominant component of the  magnetization, and the dynamics can be then be analysed in terms of spin domain coarsening, which is the focus of the next subsection.
  
\begin{figure}
\centering
\includegraphics[width=0.5\textwidth]{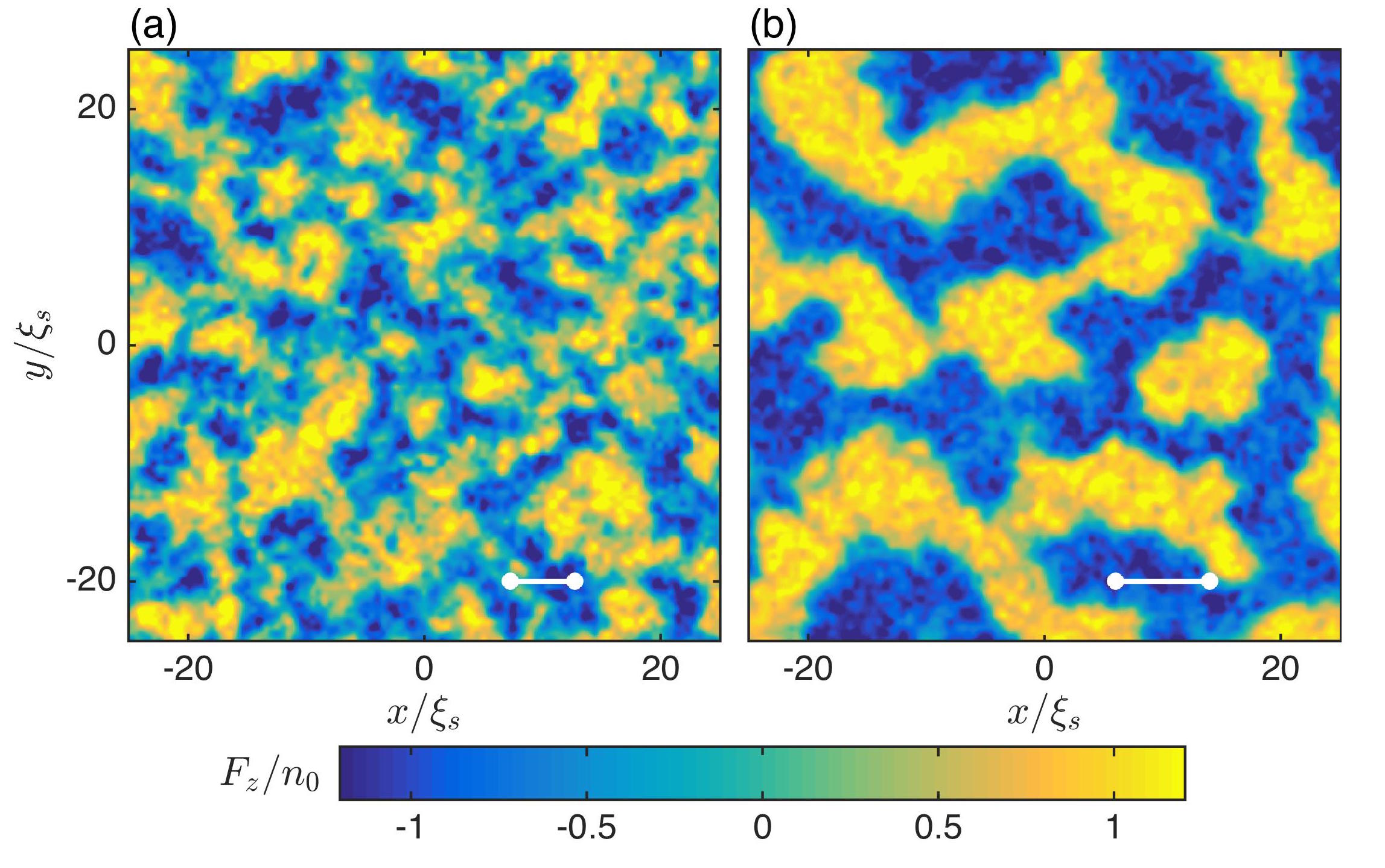}
\caption{\label{FigDomains}  Longitudinal spin density at a  time of $t=100\,t_s$ after the (a) Q1 and (b) Q2 quench to $q=-0.3q_0$. The white line  in each image indicates the spin correlation length $L_z$ defined in the text, with a value of (a) $5.5\,\xi_s$  and (b) $8.0\,\xi_s$.The spin density is only shown over a  subregion of the simulation, with other parameters as in Fig.~\ref{Figspectrum}. }
\end{figure}

We provide examples of the early-time dynamics of the local magnetization  in Fig.~\ref{Figstdynamics} for the Q1 and Q2 quenches to two different final values of $q$. We generally observe that for the Q2 quench the longitudinal magnetization  develops more rapidly and approaches a larger value at late times, compared to the Q1 case. This second observation can be understood as arising from the additional heating in the Q1 quench   (as discussed in Sec.~\ref{SecPhasesQuenches}) which tends to reduce the ground state order. Notably, for even deeper quenches ($q<-q_0$) than presented here we find that local longitudinal magnetization obtained after Q1 quench decreases as the final value of $q$ becomes more negative, while the Q2 quench is insensitive to the \ value of $q$. In Fig.~\ref{FigDomains} we compare the magnetized domains formed near the end of the early-time dynamics. These results emphasize that the Q2 quench produces larger and less noisy domains following the quench.

We can obtain qualitative insight into other aspects of the dynamics in Fig.~\ref{Figstdynamics} through considering the unstable excitations of the initial states, discussed in Sec.~\ref{Secunstablemodes}. 

 For the Q1 quench to $q=-0.3q_0$ [Fig.~\ref{Figstdynamics}(a)] and $-q_0$ [Fig.~\ref{Figstdynamics}(b)]  the dynamics proceeds in a qualitatively similar manner. The two degenerate magnon branches [Eq.~(\ref{EqMagnonP})] have unstable sections, and the growth of these excitations causes the transfer of atoms into the $m=\pm1$ sublevels (i.e.~spin-mixing), leading to the exponential growth of transverse magnetization.  After this process saturates (also see \cite{Leslie2009a}) the longitudinal magnetization then develops as the transverse magnetization decays.
 
 For  the Q2 quench the dynamics depends more strongly on the final $q$ value. For $q>-q_0$ [e.g.~Fig.~\ref{Figstdynamics}(a)] both the magnon branches [Eqs.~(\ref{EqMagnonAF1}) and (\ref{EqMagnonAF2})] are unstable. As discussed in Sec.~\ref{SecmagnonsQ2}, these magnons have transverse and longitudinal magnetic character and the simulation results show that both components of magnetization initially grow in a similar manner until saturation. For a deeper quench to $q<-q_0$, only the longitudinal magnon branch [Eq.~(\ref{EqMagnonAF1})]  is unstable, and we see [Fig.~\ref{Figstdynamics}(b)] that the longitudinal magnetization develops with negligible growth in transverse magnetization. This case has similarities to the immiscibility transition in a binary condensate \cite{Ao1998a,Hofmann2014}, in that the initial state occupies the $m=\pm1$ sublevels and the dynamics causes the condensate density in these two sublevels spatially seperate. In contrast for shallow quenches ($q>-q_0$) the unstable magnon causing transverse magnetization allows spin mixing to transfer some of the $m=\pm1$ atoms into the $m=0$ sublevel.

\subsection{Correlation functions and coarsening} \label{SecCoarsening}
In the early-time dynamics local longitudinal magnetization is established by small domains forming of   microscopic ($\sim\xi_s$) dimensions (cf.~Fig.~\ref{FigDomains}). These domains then grow by  coarsening dynamics. The average spatial distribution of domains in the system is captured by computing the relevant two-point correlation function. Here we consider the relevant spin correlation functions
\begin{align} 
G_z(\mathbf{r},t)=&\frac{1}{n_0^2l^2}\! \int\! d^2\mathbf{x}'\langle F_z(\mathbf{x}') F_z(\mathbf{x}'+\mathbf{r})\rangle_t,\label{Gz}\\
\!\!\!G_\perp(\mathbf{r},t)=&\frac{1}{n_0^2l^2}\! \int\! d^2\mathbf{x}'\langle \mathbf{F}_\perp(\mathbf{x}')\cdot \mathbf{F}_\perp(\mathbf{x}'+\mathbf{r})\rangle_t,
\end{align} 
where $\mathbf{F}_\perp\equiv(F_x,F_y)$ is the transverse magnetization density and the average is taken at a time $t$ after the quench. The functions $G_z$ and $G_\perp$ characterize longitudinal and transverse magnetization, respectively. For our case of quenches to the easy-axis phase $G_z$ is the (spin) \textit{order parameter} correlation function of interest for coarsening (e.g.~see \cite{Kudo2013a,Williamson2016a}). We also consider $G_\perp$ to demonstrate the contrasting behavior for transverse magnetisation which does not order.

Here we also address the superfluid order as a second type of order that develops in the system. This order is characterized by the global phase coherence of the atomic fields, and for the easy-axis phase this is captured by the correlation function
\begin{align} 
G_0(\mathbf{r},t)=&\frac{1}{n_0l^2}\! \int\! d^2\mathbf{x}'\langle \psi_{1}^*(\mathbf{x}') \psi_{1}(\mathbf{x}'+\mathbf{r})\rangle_t \label{G0}\\
&+\frac{1}{n_0l^2}\! \int\! d^2\mathbf{x}'\langle \psi_{-1}^*(\mathbf{x}') \psi_{-1}(\mathbf{x}'+\mathbf{r})\rangle_t.\nonumber
\end{align} 

In  expressions (\ref{Gz})-(\ref{G0}) we have utilized translational invariance to spatially average the correlation function. Additionally, we  use spatial isotropy of the correlation functions to perform an angular average over all points at a distance $r$ and further improve statistical sampling by averaging over 32 simulation trajectories conducted with different initial noise.

\begin{figure}
\centering
\includegraphics[width=0.5\textwidth]{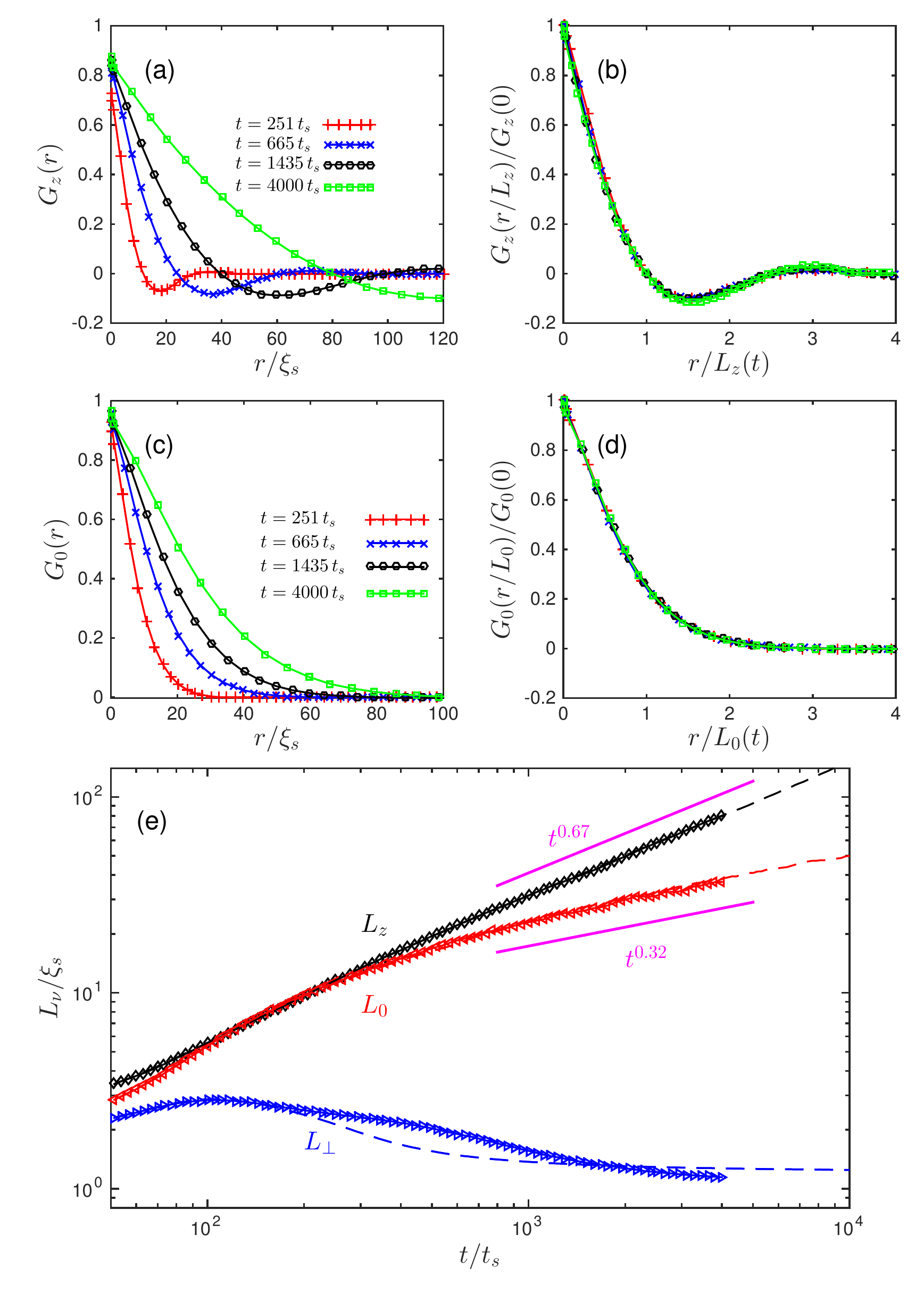}
\caption{\label{FigcorrQ1}   
Correlation functions for Q1 quench to $q=-0.3q_0$. The longitudinal spin correlation function (a) at various times and  (b) scaled by the length scale $L_z(t)$ to reveal correlation function collapse. The superfluid correlation function (c) at various times and (d) scaled by  $L_0(t)$. (e) The evolution of the length scales.  Results with symbols out to $t=4\times10^3\,t_s$ are for simulations with $l=510\,\xi_s$ and $N=1024$ points, whiles the dashed lines out to $t=10^4\,t_s$ are from simulations with $l=800\xi_s$ with $N=1024$ points.
}
\end{figure}

\begin{figure}
\centering
\includegraphics[width=0.5\textwidth]{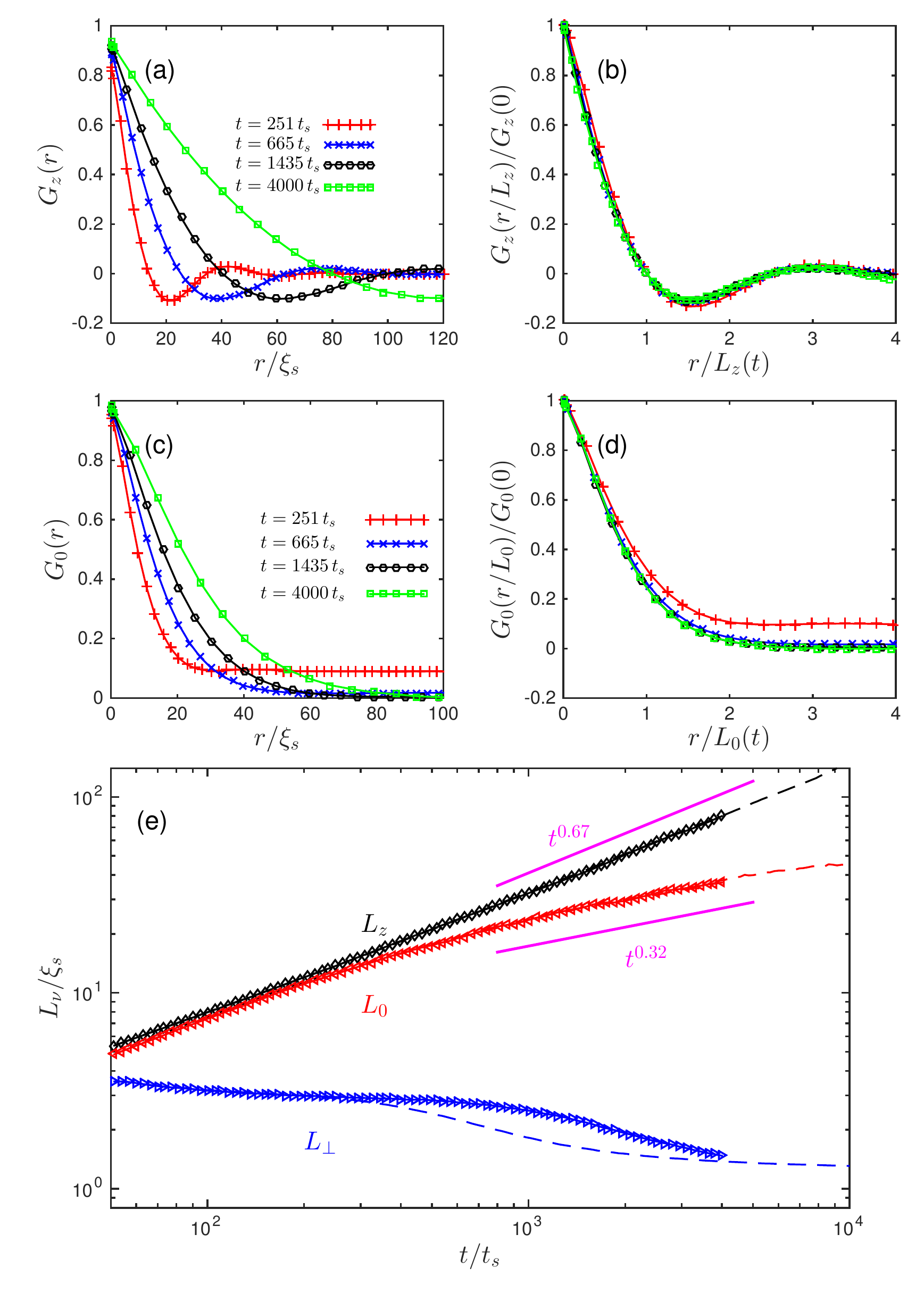}
\caption{\label{FigcorrQ2}   
Correlation functions for Q2 quench to $q=-0.3q_0$.  The longitudinal spin correlation function (a) at various times and  (b) scaled by the length scale $L_z(t)$ to reveal correlation function collapse. The superfluid correlation function (c) at various times and (d) scaled by  $L_0(t)$. (e) The evolution of the length scales.  Results with symbols out to $t=4\times10^3\,t_s$ are for simulations with $l=510\,\xi_s$ and $N=1024$ points, whiles the dashed lines out to $t=10^4\,t_s$ are from simulations with $l=800\xi_s$ with $N=1024$ points.
}
\end{figure}

The temporal evolution of the $G_z$ and $G_0$ correlation functions is shown in  Figs.~\ref{FigcorrQ1} and \ref{FigcorrQ2} for both types of quench.  The length scale over which the correlation function decays can be taken to define a characteristic domain size.  For $G_z$ we take this length scale $L_z(t)$ to be the first zero crossing of $G_z(r,t)$. The $G_0$ and $G_\perp$ (not shown) correlation functions do not usually have a  zero crossing so we take $L_0(t)$ to be the position where the correlation peak in $G_0$ decays to 0.25 of its maximal value (occurring at $r=0$), and using the same procedure we define the length scale $L_\perp$ for $G_\perp$.  

As time progresses the length scales $L_z$ and $L_0$ are seen to grow as the longitudinal spin order and superfluid order extends over large regions [Figs.~\ref{FigcorrQ1}(e) and  \ref{FigcorrQ2}(e)]. In contrast the length scale $L_\perp$, which is not associated with the ground state order of the system, does not grow and remains comparable to the microscopic length scale $\xi_s$.

We investigate whether the system exhibits dynamic scale invariance, i.e.~whether correlations of the order parameters at late times collapse onto a single universal curve $H_\nu(r)$ when lengths are scaled by a  characteristic length scale $L_\nu(t)$ 
\begin{align}
H_\nu(r)=G_\nu(r/L_\nu(t),t),\qquad \nu=z,0.
\end{align} 
Using the length scales $L_z(t)$ and $L_0(t)$ we demonstrate  correlation function collapse for sufficiently late times in both the spin order  [see Figs.~\ref{FigcorrQ1}(b),\ref{FigcorrQ2}(b)] and the superfluid order [see Figs.~\ref{FigcorrQ1}(d), \ref{FigcorrQ2}(d)].

For both quenches at late times ($t\gtrsim10^2t_s$) $L_z$ is seen to grow as $L_z\sim t^{2/3}$ as found previously \cite{Kudo2013a,Williamson2016a}. 
The superfluid order (length scale $L_0$) grows in a similar manner to $L_z$ for intermediate times, but eventually ($t\gtrsim5\times10^2\,t_s$) grows more slowly. Our best fit to the exponent gives $z\sim3.2\pm0.2$.
This result is slower than the $L_0(t)\sim(t/\ln t)^{1/2}$ scaling expected from dissipative XY-model dynamics \cite{Yurke1993a,Bray2000a}, where the logarithmic correction to the $z=2$ exponent originates from the presence of free vortices. For the case of a quasi-2D scalar condensate  Damle \textit{et al.}~\cite{Damle1996a} found $z\approx1.1$, albeit for a higher temperature regime (close to the critical temperature for vortex unbinding) and using small simulation grids. Recent results of Karl \textit{et al.}~\cite{Karl2016a} reveal even slower dynamics ($z\approx5$) for particular initial condition of a quasi-2D condensate. We have performed simulations for deeper quenches to $q=-q_0$ and for the case of a larger interaction ratio $g_s=-g_n/12$ to verify that the spin and superfluid ordering occurs with the same exponents.

We noted in Sec.~\ref{SecEarlyTime} that local order developed faster in the early-time dynamics for the Q2 quench relative to Q1. Comparing Figs.~\ref{FigcorrQ1}(e) and \ref{FigcorrQ2}(e) we also observe that for times $t\lesssim3\times10^2\,t_s$ the values of $L_z$ and $L_0$ for the Q2 quench are appreciably larger that those of the Q1 quench at the same time, and grow at a more steady rate [the $L_z$ correlation lengths at $t=10^2\,t_s$ are also shown in Fig.~\ref{FigDomains}]. The $G_0$ correlation function at $t=251\,t_s$ in Fig.~\ref{FigcorrQ2}(c) shows that the system has long-range superfluid order at early times (i.e.~$G_0$ approaches a constant non-zero value for large $r$). This order eventually decays ($t\gtrsim5\times10^2\,t_s$) and the $G_0$ behavior is similar to that observed in the Q1 quench.
 The long-range order is established  by the initial spin rotation used in Q2 quench, which rotates the coherent polar state into $m\pm1$ sublevels. In the postquench dynamics, this coherence is rapidly destroyed by the immiscibility dynamics leading to spin domain formation. 

In Figs.~\ref{FigcorrQ1}(e) and \ref{FigcorrQ2}(e) we also show results for a calculation with a grid of $N=1024$ and $l=800\xi_s$ (i.e.~with a grid point spacing about 60\% larger than the other results). The greater spatial range in this case allows us to simulate longer (we need to ensure that $L_z,L_0\ll l$ to avoid finite size effects). The results from these calculations are shown as dashed lines and are seen to lie on top of the main results for characteristic length scales associated with the order parameter (i.e.~ $L_z$ and $L_0$), but differ appreciably from the main results for the behavior of $L_\perp$. This nicely demonstrates the universality of the order parameter coarsening dynamics, while emphasizing that the transverse magnetization is a disordered degree of freedom that is sensitive to the microscopic details (in this case it is largely determined by the thermalization of the $m=0$ spin waves, see \cite{Williamson2016b}).

\subsection{Domain Size Distribution} 
The binary character of the easy-axis spin order allows clear identification of spin domains (e.g.~see Fig.~\ref{FigDomains}). In this subsection we consider the properties of the domains produced as characterized by the domain area distribution $\rho(S,t)$ where $\rho(S,t)dS$ is the average number of domains of area $S$ to $S+dS$ per unit system area at time $t$. In simulations we define positive domains as connected spatial regions where the $F_z$ spin density is positive\footnote{More precisely we require that $F_z\ge0.1n_0$ to be part of a positive domain.}, and all locations satisfying this are set to a value of +1, and all other locations are set to a value of zero. On this \textit{binary image} we apply the Hoshen-Kopelman algorithm \cite{Hoshen1976a} to label the clusters of contiguous +1 cells. Similarly, a binary image can be constructed for the negative domains (with $F_z<0$). Analyzing the combined set of positive and negative magnetized domains at time $t$ we compute their areas, and bin this to determine $\rho(S,t)$. To improve statistics we use 32 trajectories to average for  $\rho(S,t)$.
 
  \begin{figure}
\centering
\includegraphics[width=0.5\textwidth]{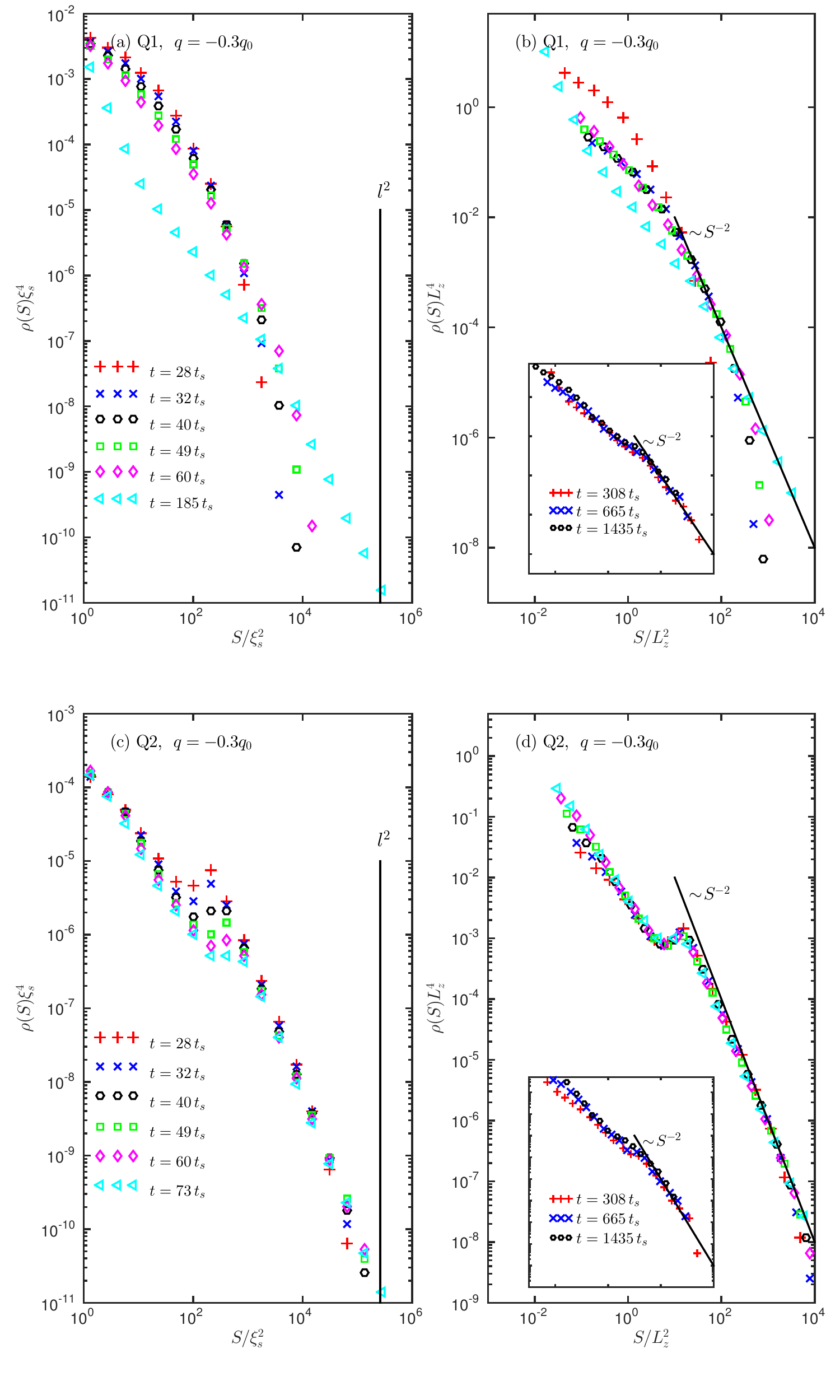}
\caption{\label{FigDomainAreas}
Domain area distributions for the (a)-(b) Q1 and (c)-(d) Q2 quenches.  The domains area distribution is shown at early-times as a function of (a), (c) area and of (b), (d) scaled-area $S/L_z^2$, where $L_z$ is the longitudinal spin correlation length evaluated as discussed in Sec.~\ref{SecCoarsening}. The insets in (b) and (d) use the same axis range as the main plots, and indicate results at later times for comparison. }
\end{figure}
 
Our results for $\rho(S,t)$  at various times are shown in Fig.~\ref{FigDomainAreas}. While the late time behavior of the domain distributions is reasonably similar [see insets to Figs.~\ref{FigDomainAreas}(b) and (d)], the early-time behavior is strikingly different. Notably, even for times as short as $t\sim40\,t_s$ in the Q2 case , $\rho(S,t)$ is seen to extend out to areas of order the system size   [i.e.~$S\sim l^2$, see Fig.~\ref{FigDomainAreas}(c)] , which indicates that the system can have a percolating domain that connects across the system.  In contrast, such percolating domains for the Q1 quench only develop on long time scales [i.e.~$t\sim200\,t_s$ in Fig.~\ref{FigDomainAreas}(a)]. This difference likely originates from the nature of the  $\epsilon_1$ magnon branch for the Q2 quench which has unstable wavelengths extending from $\sim\xi_s$ up to the system size [e.g.~see Fig.~\ref{Figspectrum}(c)]. In comparison the unstable modes for the Q1 quench are restricted to a finite wavelength range around $\lambda\sim\sqrt{-q_0/2q}\,\xi_s$ [e.g.~see Fig.~\ref{Figspectrum}(a)].

We can apply dynamic scaling to the domain size distribution. Since the $G_z$ correlation function grows as $L_z(t)$,  that the domain areas should grow as $L_z(t)^2$, and if dynamic scaling holds we would expect 
\begin{equation}
\tilde{\rho}\equiv\rho(S/L_z(t)^2,t)L_z(t)^4,
\end{equation}
to be a universal time-independent distribution. We perform this scaling in Figs.~\ref{FigDomainAreas}(b) and (d). At late-times (insets to those figures) we see reasonable collapse of the area distribution, however at earlier-times ($t\lesssim100\,t_s$) only the Q2 quench exhibits scaling collapse over a wide range of areas. We also indicate on the scaled results the $S^{-2}$ power law, which provides a reasonable fit to the distribution for sufficiently large domains. Such power law behavior in domain size (or ``cluster size") distribution is predicted in percolation theory near the percolation transition with $\tau\approx2$  in 2D. Normally in percolation theory the occupation probability ($p$) is varied to pass through the transition. For the spin-1 system this is controlled by the total (conserved) $z$-magnetization, and for our results (where the total $z$-magnetization is zero) the probability of a location being in a positive (or negative) domain is $p=0.5$, which is approximately the percolation threshold condition in 2D.
 In order to explore the percolation transition it would be necessary to vary the total $z$-magnetization, which would in turn vary the relative portion of two domain types. Such a study has been performed using simulations of a segregating binary condensate, where the effective magnetization is controlled by varying the relative portion of the two atomic species \cite{Takeuchi2015a}. This study presented strong evidence for a percolation threshold at $p\approx0.5$.

\section{Conclusion}\label{SecConclude}
We have considered the spin and superfluid ordering of a spin-1 condensate. We have found that in the coarsening dynamics these two types of order exhibit dynamic scaling, but develop with different dynamic critical exponents. We find that (and as previously observed in Refs.~\cite{Kudo2013a,Williamson2016a,Williamson2016b}) the spin-order develops with a growth law of $L_z(t) \sim t^{2/3}$, consistent with a binary fluid in the inertial hydrodynamic regime \cite{Furukawa1985}. Our new results here concern the superfluid ordering, which we find to grow much more slowly as $L_0(t)\sim t^{0.32}$. Indeed, our simulations out to $t=10^4\,t_s$ reveal that $L_0$ is almost an order of magnitude smaller than $L_z$ [see Figs.~\ref{FigcorrQ1}(e) and \ref{FigcorrQ2}(e)]. Future work will involve developing an understanding for the slower superfluid growth rate, and whether aspects of the spin domains (e.g.~shedding of various types of vortices at the spin domain walls, or other types of topological defects) play a role. Another direction for investigation is to consider our observed dynamics in the context of pre-thermalization and turbulence theory, which have proven fruitful directions of research for the immiscibility transition in a binary condensate \cite{Karl2013a} (also see \cite{Fujimoto2016a}).

In addition to the standard Q1 quench, in which the quadratic Zeeman energy is suddenly changed, we propose and simulate a second type of quench (Q2) in which a spin rotation is applied prior to the Zeeman energy being changed. In the late-time dynamics both quenches exhibit the same ordering dynamics, demonstrating the universal nature of the coarsening dynamics. However, the Q2 quench has favourable properties for exploring domain formation: (i) the domains form more rapidly and with less heating; (ii) the domains percolate at early times. These properties may make the Q2 quench better suited to experiments (e.g.,~with $^{87}$Rb condensates), where the small spin-dependent interaction makes it challenging to explore long-time post-quench dynamics. For example, some of the short-time predictions we have made for domain formation and the behavior of the spin domain area distribution should be feasible to study in current experiments.

\section*{Acknowledgments}
The authors acknowledge useful discussions and assistance from H.~Takeuchi, L.~Williamson, and L.~Symes.  
 PBB acknowledges the contribution of NZ eScience Infrastructure (NeSI) high-performance computing facilities, and support from the Marsden Fund of the Royal Society of New Zealand.   
 

\begin{thebibliography}{50}%
\makeatletter
\providecommand \@ifxundefined [1]{%
 \@ifx{#1\undefined}
}%
\providecommand \@ifnum [1]{%
 \ifnum #1\expandafter \@firstoftwo
 \else \expandafter \@secondoftwo
 \fi
}%
\providecommand \@ifx [1]{%
 \ifx #1\expandafter \@firstoftwo
 \else \expandafter \@secondoftwo
 \fi
}%
\providecommand \natexlab [1]{#1}%
\providecommand \enquote  [1]{``#1''}%
\providecommand \bibnamefont  [1]{#1}%
\providecommand \bibfnamefont [1]{#1}%
\providecommand \citenamefont [1]{#1}%
\providecommand \href@noop [0]{\@secondoftwo}%
\providecommand \href [0]{\begingroup \@sanitize@url \@href}%
\providecommand \@href[1]{\@@startlink{#1}\@@href}%
\providecommand \@@href[1]{\endgroup#1\@@endlink}%
\providecommand \@sanitize@url [0]{\catcode `\\12\catcode `\$12\catcode
  `\&12\catcode `\#12\catcode `\^12\catcode `\_12\catcode `\%12\relax}%
\providecommand \@@startlink[1]{}%
\providecommand \@@endlink[0]{}%
\providecommand \url  [0]{\begingroup\@sanitize@url \@url }%
\providecommand \@url [1]{\endgroup\@href {#1}{\urlprefix }}%
\providecommand \urlprefix  [0]{URL }%
\providecommand \Eprint [0]{\href }%
\providecommand \doibase [0]{http://dx.doi.org/}%
\providecommand \selectlanguage [0]{\@gobble}%
\providecommand \bibinfo  [0]{\@secondoftwo}%
\providecommand \bibfield  [0]{\@secondoftwo}%
\providecommand \translation [1]{[#1]}%
\providecommand \BibitemOpen [0]{}%
\providecommand \bibitemStop [0]{}%
\providecommand \bibitemNoStop [0]{.\EOS\space}%
\providecommand \EOS [0]{\spacefactor3000\relax}%
\providecommand \BibitemShut  [1]{\csname bibitem#1\endcsname}%
\let\auto@bib@innerbib\@empty
\bibitem [{\citenamefont {Kawaguchi}\ and\ \citenamefont
  {Ueda}(2012)}]{Kawaguchi2012R}%
  \BibitemOpen
  \bibfield  {author} {\bibinfo {author} {\bibfnamefont {Y.}~\bibnamefont
  {Kawaguchi}}\ and\ \bibinfo {author} {\bibfnamefont {M.}~\bibnamefont
  {Ueda}},\ }\href {\doibase http://dx.doi.org/10.1016/j.physrep.2012.07.005}
  {\bibfield  {journal} {\bibinfo  {journal} {Physics Reports}\ }\textbf
  {\bibinfo {volume} {520}},\ \bibinfo {pages} {253 } (\bibinfo {year}
  {2012})}\BibitemShut {NoStop}%
\bibitem [{\citenamefont {Stamper-Kurn}\ and\ \citenamefont
  {Ueda}(2013)}]{StamperKurn2013a}%
  \BibitemOpen
  \bibfield  {author} {\bibinfo {author} {\bibfnamefont {D.~M.}\ \bibnamefont
  {Stamper-Kurn}}\ and\ \bibinfo {author} {\bibfnamefont {M.}~\bibnamefont
  {Ueda}},\ }\href {\doibase 10.1103/RevModPhys.85.1191} {\bibfield  {journal}
  {\bibinfo  {journal} {Rev. Mod. Phys.}\ }\textbf {\bibinfo {volume} {85}},\
  \bibinfo {pages} {1191} (\bibinfo {year} {2013})}\BibitemShut {NoStop}%
\bibitem [{\citenamefont {Sadler}\ \emph {et~al.}(2006)\citenamefont {Sadler},
  \citenamefont {Higbie}, \citenamefont {Leslie}, \citenamefont
  {Vengalattore},\ and\ \citenamefont {Stamper-Kurn}}]{Sadler2006a}%
  \BibitemOpen
  \bibfield  {author} {\bibinfo {author} {\bibfnamefont {L.~E.}\ \bibnamefont
  {Sadler}}, \bibinfo {author} {\bibfnamefont {J.~M.}\ \bibnamefont {Higbie}},
  \bibinfo {author} {\bibfnamefont {S.~R.}\ \bibnamefont {Leslie}}, \bibinfo
  {author} {\bibfnamefont {M.}~\bibnamefont {Vengalattore}}, \ and\ \bibinfo
  {author} {\bibfnamefont {D.~M.}\ \bibnamefont {Stamper-Kurn}},\ }\href
  {http://dx.doi.org/10.1038/nature05094} {\bibfield  {journal} {\bibinfo
  {journal} {Nature}\ }\textbf {\bibinfo {volume} {443}},\ \bibinfo {pages}
  {312} (\bibinfo {year} {2006})}\BibitemShut {NoStop}%
\bibitem [{\citenamefont {Leslie}\ \emph {et~al.}(2009)\citenamefont {Leslie},
  \citenamefont {Guzman}, \citenamefont {Vengalattore}, \citenamefont {Sau},
  \citenamefont {Cohen},\ and\ \citenamefont {Stamper-Kurn}}]{Leslie2009a}%
  \BibitemOpen
  \bibfield  {author} {\bibinfo {author} {\bibfnamefont {S.~R.}\ \bibnamefont
  {Leslie}}, \bibinfo {author} {\bibfnamefont {J.}~\bibnamefont {Guzman}},
  \bibinfo {author} {\bibfnamefont {M.}~\bibnamefont {Vengalattore}}, \bibinfo
  {author} {\bibfnamefont {J.~D.}\ \bibnamefont {Sau}}, \bibinfo {author}
  {\bibfnamefont {M.~L.}\ \bibnamefont {Cohen}}, \ and\ \bibinfo {author}
  {\bibfnamefont {D.~M.}\ \bibnamefont {Stamper-Kurn}},\ }\href {\doibase
  10.1103/PhysRevA.79.043631} {\bibfield  {journal} {\bibinfo  {journal} {Phys.
  Rev. A}\ }\textbf {\bibinfo {volume} {79}},\ \bibinfo {pages} {043631}
  (\bibinfo {year} {2009})}\BibitemShut {NoStop}%
\bibitem [{\citenamefont {Bookjans}\ \emph {et~al.}(2011)\citenamefont
  {Bookjans}, \citenamefont {Vinit},\ and\ \citenamefont
  {Raman}}]{Bookjans2011b}%
  \BibitemOpen
  \bibfield  {author} {\bibinfo {author} {\bibfnamefont {E.~M.}\ \bibnamefont
  {Bookjans}}, \bibinfo {author} {\bibfnamefont {A.}~\bibnamefont {Vinit}}, \
  and\ \bibinfo {author} {\bibfnamefont {C.}~\bibnamefont {Raman}},\ }\href
  {\doibase 10.1103/PhysRevLett.107.195306} {\bibfield  {journal} {\bibinfo
  {journal} {Phys. Rev. Lett.}\ }\textbf {\bibinfo {volume} {107}},\ \bibinfo
  {pages} {195306} (\bibinfo {year} {2011})}\BibitemShut {NoStop}%
\bibitem [{\citenamefont {Guzman}\ \emph {et~al.}(2011)\citenamefont {Guzman},
  \citenamefont {Jo}, \citenamefont {Wenz}, \citenamefont {Murch},
  \citenamefont {Thomas},\ and\ \citenamefont {Stamper-Kurn}}]{Guzman2011a}%
  \BibitemOpen
  \bibfield  {author} {\bibinfo {author} {\bibfnamefont {J.}~\bibnamefont
  {Guzman}}, \bibinfo {author} {\bibfnamefont {G.-B.}\ \bibnamefont {Jo}},
  \bibinfo {author} {\bibfnamefont {A.~N.}\ \bibnamefont {Wenz}}, \bibinfo
  {author} {\bibfnamefont {K.~W.}\ \bibnamefont {Murch}}, \bibinfo {author}
  {\bibfnamefont {C.~K.}\ \bibnamefont {Thomas}}, \ and\ \bibinfo {author}
  {\bibfnamefont {D.~M.}\ \bibnamefont {Stamper-Kurn}},\ }\href {\doibase
  10.1103/PhysRevA.84.063625} {\bibfield  {journal} {\bibinfo  {journal} {Phys.
  Rev. A}\ }\textbf {\bibinfo {volume} {84}},\ \bibinfo {pages} {063625}
  (\bibinfo {year} {2011})}\BibitemShut {NoStop}%
\bibitem [{\citenamefont {De}\ \emph {et~al.}(2014)\citenamefont {De},
  \citenamefont {Campbell}, \citenamefont {Price}, \citenamefont {Putra},
  \citenamefont {Anderson},\ and\ \citenamefont {Spielman}}]{De2014a}%
  \BibitemOpen
  \bibfield  {author} {\bibinfo {author} {\bibfnamefont {S.}~\bibnamefont
  {De}}, \bibinfo {author} {\bibfnamefont {D.~L.}\ \bibnamefont {Campbell}},
  \bibinfo {author} {\bibfnamefont {R.~M.}\ \bibnamefont {Price}}, \bibinfo
  {author} {\bibfnamefont {A.}~\bibnamefont {Putra}}, \bibinfo {author}
  {\bibfnamefont {B.~M.}\ \bibnamefont {Anderson}}, \ and\ \bibinfo {author}
  {\bibfnamefont {I.~B.}\ \bibnamefont {Spielman}},\ }\href {\doibase
  10.1103/PhysRevA.89.033631} {\bibfield  {journal} {\bibinfo  {journal} {Phys.
  Rev. A}\ }\textbf {\bibinfo {volume} {89}},\ \bibinfo {pages} {033631}
  (\bibinfo {year} {2014})}\BibitemShut {NoStop}%
\bibitem [{\citenamefont {Lamacraft}(2007)}]{Lamacraft2007a}%
  \BibitemOpen
  \bibfield  {author} {\bibinfo {author} {\bibfnamefont {A.}~\bibnamefont
  {Lamacraft}},\ }\href {\doibase 10.1103/PhysRevLett.98.160404} {\bibfield
  {journal} {\bibinfo  {journal} {Phys. Rev. Lett.}\ }\textbf {\bibinfo
  {volume} {98}},\ \bibinfo {pages} {160404} (\bibinfo {year}
  {2007})}\BibitemShut {NoStop}%
\bibitem [{\citenamefont {Saito}\ \emph
  {et~al.}(2007{\natexlab{a}})\citenamefont {Saito}, \citenamefont
  {Kawaguchi},\ and\ \citenamefont {Ueda}}]{Saito2007a}%
  \BibitemOpen
  \bibfield  {author} {\bibinfo {author} {\bibfnamefont {H.}~\bibnamefont
  {Saito}}, \bibinfo {author} {\bibfnamefont {Y.}~\bibnamefont {Kawaguchi}}, \
  and\ \bibinfo {author} {\bibfnamefont {M.}~\bibnamefont {Ueda}},\ }\href
  {\doibase 10.1103/PhysRevA.76.043613} {\bibfield  {journal} {\bibinfo
  {journal} {Phys. Rev. A}\ }\textbf {\bibinfo {volume} {76}},\ \bibinfo
  {pages} {043613} (\bibinfo {year} {2007}{\natexlab{a}})}\BibitemShut
  {NoStop}%
\bibitem [{\citenamefont {Saito}\ \emph
  {et~al.}(2007{\natexlab{b}})\citenamefont {Saito}, \citenamefont
  {Kawaguchi},\ and\ \citenamefont {Ueda}}]{Saito2007b}%
  \BibitemOpen
  \bibfield  {author} {\bibinfo {author} {\bibfnamefont {H.}~\bibnamefont
  {Saito}}, \bibinfo {author} {\bibfnamefont {Y.}~\bibnamefont {Kawaguchi}}, \
  and\ \bibinfo {author} {\bibfnamefont {M.}~\bibnamefont {Ueda}},\ }\href
  {\doibase 10.1103/PhysRevA.75.013621} {\bibfield  {journal} {\bibinfo
  {journal} {Phys. Rev. A}\ }\textbf {\bibinfo {volume} {75}},\ \bibinfo
  {pages} {013621} (\bibinfo {year} {2007}{\natexlab{b}})}\BibitemShut
  {NoStop}%
\bibitem [{\citenamefont {Uhlmann}\ \emph {et~al.}(2007)\citenamefont
  {Uhlmann}, \citenamefont {Sch\"utzhold},\ and\ \citenamefont
  {Fischer}}]{Uhlmann2007a}%
  \BibitemOpen
  \bibfield  {author} {\bibinfo {author} {\bibfnamefont {M.}~\bibnamefont
  {Uhlmann}}, \bibinfo {author} {\bibfnamefont {R.}~\bibnamefont
  {Sch\"utzhold}}, \ and\ \bibinfo {author} {\bibfnamefont {U.~R.}\
  \bibnamefont {Fischer}},\ }\href {\doibase 10.1103/PhysRevLett.99.120407}
  {\bibfield  {journal} {\bibinfo  {journal} {Phys. Rev. Lett.}\ }\textbf
  {\bibinfo {volume} {99}},\ \bibinfo {pages} {120407} (\bibinfo {year}
  {2007})}\BibitemShut {NoStop}%
\bibitem [{\citenamefont {Damski}\ and\ \citenamefont
  {Zurek}(2007)}]{Damski2007a}%
  \BibitemOpen
  \bibfield  {author} {\bibinfo {author} {\bibfnamefont {B.}~\bibnamefont
  {Damski}}\ and\ \bibinfo {author} {\bibfnamefont {W.~H.}\ \bibnamefont
  {Zurek}},\ }\href {\doibase 10.1103/PhysRevLett.99.130402} {\bibfield
  {journal} {\bibinfo  {journal} {Phys. Rev. Lett.}\ }\textbf {\bibinfo
  {volume} {99}},\ \bibinfo {pages} {130402} (\bibinfo {year}
  {2007})}\BibitemShut {NoStop}%
\bibitem [{\citenamefont {Vinit}\ \emph {et~al.}(2013)\citenamefont {Vinit},
  \citenamefont {Bookjans}, \citenamefont {S\'a~de Melo},\ and\ \citenamefont
  {Raman}}]{Vinit2013a}%
  \BibitemOpen
  \bibfield  {author} {\bibinfo {author} {\bibfnamefont {A.}~\bibnamefont
  {Vinit}}, \bibinfo {author} {\bibfnamefont {E.~M.}\ \bibnamefont {Bookjans}},
  \bibinfo {author} {\bibfnamefont {C.~A.~R.}\ \bibnamefont {S\'a~de Melo}}, \
  and\ \bibinfo {author} {\bibfnamefont {C.}~\bibnamefont {Raman}},\ }\href
  {\doibase 10.1103/PhysRevLett.110.165301} {\bibfield  {journal} {\bibinfo
  {journal} {Phys. Rev. Lett.}\ }\textbf {\bibinfo {volume} {110}},\ \bibinfo
  {pages} {165301} (\bibinfo {year} {2013})}\BibitemShut {NoStop}%
\bibitem [{\citenamefont {Witkowska}\ \emph {et~al.}(2014)\citenamefont
  {Witkowska}, \citenamefont {\ifmmode~\acute{S}\else \'{S}\fi{}wis\l{}ocki},\
  and\ \citenamefont {Matuszewski}}]{Witkowska2014a}%
  \BibitemOpen
  \bibfield  {author} {\bibinfo {author} {\bibfnamefont {E.}~\bibnamefont
  {Witkowska}}, \bibinfo {author} {\bibfnamefont {T.}~\bibnamefont
  {\ifmmode~\acute{S}\else \'{S}\fi{}wis\l{}ocki}}, \ and\ \bibinfo {author}
  {\bibfnamefont {M.}~\bibnamefont {Matuszewski}},\ }\href {\doibase
  10.1103/PhysRevA.90.033604} {\bibfield  {journal} {\bibinfo  {journal} {Phys.
  Rev. A}\ }\textbf {\bibinfo {volume} {90}},\ \bibinfo {pages} {033604}
  (\bibinfo {year} {2014})}\BibitemShut {NoStop}%
\bibitem [{\citenamefont {Mukerjee}\ \emph {et~al.}(2007)\citenamefont
  {Mukerjee}, \citenamefont {Xu},\ and\ \citenamefont {Moore}}]{Mukerjee2007a}%
  \BibitemOpen
  \bibfield  {author} {\bibinfo {author} {\bibfnamefont {S.}~\bibnamefont
  {Mukerjee}}, \bibinfo {author} {\bibfnamefont {C.}~\bibnamefont {Xu}}, \ and\
  \bibinfo {author} {\bibfnamefont {J.~E.}\ \bibnamefont {Moore}},\ }\href
  {\doibase 10.1103/PhysRevB.76.104519} {\bibfield  {journal} {\bibinfo
  {journal} {Phys. Rev. B}\ }\textbf {\bibinfo {volume} {76}},\ \bibinfo
  {pages} {104519} (\bibinfo {year} {2007})}\BibitemShut {NoStop}%
\bibitem [{\citenamefont {Vengalattore}\ \emph {et~al.}(2010)\citenamefont
  {Vengalattore}, \citenamefont {Guzman}, \citenamefont {Leslie}, \citenamefont
  {Serwane},\ and\ \citenamefont {Stamper-Kurn}}]{Vengalattore2010a}%
  \BibitemOpen
  \bibfield  {author} {\bibinfo {author} {\bibfnamefont {M.}~\bibnamefont
  {Vengalattore}}, \bibinfo {author} {\bibfnamefont {J.}~\bibnamefont
  {Guzman}}, \bibinfo {author} {\bibfnamefont {S.~R.}\ \bibnamefont {Leslie}},
  \bibinfo {author} {\bibfnamefont {F.}~\bibnamefont {Serwane}}, \ and\
  \bibinfo {author} {\bibfnamefont {D.~M.}\ \bibnamefont {Stamper-Kurn}},\
  }\href {\doibase 10.1103/PhysRevA.81.053612} {\bibfield  {journal} {\bibinfo
  {journal} {Phys. Rev. A}\ }\textbf {\bibinfo {volume} {81}},\ \bibinfo
  {pages} {053612} (\bibinfo {year} {2010})}\BibitemShut {NoStop}%
\bibitem [{\citenamefont {Kudo}\ and\ \citenamefont
  {Kawaguchi}(2013)}]{Kudo2013a}%
  \BibitemOpen
  \bibfield  {author} {\bibinfo {author} {\bibfnamefont {K.}~\bibnamefont
  {Kudo}}\ and\ \bibinfo {author} {\bibfnamefont {Y.}~\bibnamefont
  {Kawaguchi}},\ }\href {\doibase 10.1103/PhysRevA.88.013630} {\bibfield
  {journal} {\bibinfo  {journal} {Phys. Rev. A}\ }\textbf {\bibinfo {volume}
  {88}},\ \bibinfo {pages} {013630} (\bibinfo {year} {2013})}\BibitemShut
  {NoStop}%
\bibitem [{\citenamefont {Kudo}\ and\ \citenamefont
  {Kawaguchi}(2015)}]{Kudo2015a}%
  \BibitemOpen
  \bibfield  {author} {\bibinfo {author} {\bibfnamefont {K.}~\bibnamefont
  {Kudo}}\ and\ \bibinfo {author} {\bibfnamefont {Y.}~\bibnamefont
  {Kawaguchi}},\ }\href {\doibase 10.1103/PhysRevA.91.053609} {\bibfield
  {journal} {\bibinfo  {journal} {Phys. Rev. A}\ }\textbf {\bibinfo {volume}
  {91}},\ \bibinfo {pages} {053609} (\bibinfo {year} {2015})}\BibitemShut
  {NoStop}%
\bibitem [{\citenamefont {Williamson}\ and\ \citenamefont
  {Blakie}(2016{\natexlab{a}})}]{Williamson2016a}%
  \BibitemOpen
  \bibfield  {author} {\bibinfo {author} {\bibfnamefont {L.~A.}\ \bibnamefont
  {Williamson}}\ and\ \bibinfo {author} {\bibfnamefont {P.~B.}\ \bibnamefont
  {Blakie}},\ }\href {\doibase 10.1103/PhysRevLett.116.025301} {\bibfield
  {journal} {\bibinfo  {journal} {Phys. Rev. Lett.}\ }\textbf {\bibinfo
  {volume} {116}},\ \bibinfo {pages} {025301} (\bibinfo {year}
  {2016}{\natexlab{a}})}\BibitemShut {NoStop}%
\bibitem [{\citenamefont {Williamson}\ and\ \citenamefont
  {Blakie}(2016{\natexlab{b}})}]{Williamson2016b}%
  \BibitemOpen
  \bibfield  {author} {\bibinfo {author} {\bibfnamefont {L.~A.}\ \bibnamefont
  {Williamson}}\ and\ \bibinfo {author} {\bibfnamefont {P.~B.}\ \bibnamefont
  {Blakie}},\ }\href {\doibase 10.1103/PhysRevA.94.023608} {\bibfield
  {journal} {\bibinfo  {journal} {Phys. Rev. A}\ }\textbf {\bibinfo {volume}
  {94}},\ \bibinfo {pages} {023608} (\bibinfo {year}
  {2016}{\natexlab{b}})}\BibitemShut {NoStop}%
\bibitem [{\citenamefont {Hofmann}\ \emph {et~al.}(2014)\citenamefont
  {Hofmann}, \citenamefont {Natu},\ and\ \citenamefont
  {Das~Sarma}}]{Hofmann2014}%
  \BibitemOpen
  \bibfield  {author} {\bibinfo {author} {\bibfnamefont {J.}~\bibnamefont
  {Hofmann}}, \bibinfo {author} {\bibfnamefont {S.~S.}\ \bibnamefont {Natu}}, \
  and\ \bibinfo {author} {\bibfnamefont {S.}~\bibnamefont {Das~Sarma}},\ }\href
  {\doibase 10.1103/PhysRevLett.113.095702} {\bibfield  {journal} {\bibinfo
  {journal} {Phys. Rev. Lett.}\ }\textbf {\bibinfo {volume} {113}},\ \bibinfo
  {pages} {095702} (\bibinfo {year} {2014})}\BibitemShut {NoStop}%
\bibitem [{\citenamefont {Karl}\ \emph {et~al.}(2013)\citenamefont {Karl},
  \citenamefont {Nowak},\ and\ \citenamefont {Gasenzer}}]{Karl2013a}%
  \BibitemOpen
  \bibfield  {author} {\bibinfo {author} {\bibfnamefont {M.}~\bibnamefont
  {Karl}}, \bibinfo {author} {\bibfnamefont {B.}~\bibnamefont {Nowak}}, \ and\
  \bibinfo {author} {\bibfnamefont {T.}~\bibnamefont {Gasenzer}},\ }\href
  {\doibase 10.1103/PhysRevA.88.063615} {\bibfield  {journal} {\bibinfo
  {journal} {Phys. Rev. A}\ }\textbf {\bibinfo {volume} {88}},\ \bibinfo
  {pages} {063615} (\bibinfo {year} {2013})}\BibitemShut {NoStop}%
\bibitem [{\citenamefont {Higbie}\ \emph {et~al.}(2005)\citenamefont {Higbie},
  \citenamefont {Sadler}, \citenamefont {Inouye}, \citenamefont {Chikkatur},
  \citenamefont {Leslie}, \citenamefont {Moore}, \citenamefont {Savalli},\ and\
  \citenamefont {Stamper-Kurn}}]{Higbie2005a}%
  \BibitemOpen
  \bibfield  {author} {\bibinfo {author} {\bibfnamefont {J.~M.}\ \bibnamefont
  {Higbie}}, \bibinfo {author} {\bibfnamefont {L.~E.}\ \bibnamefont {Sadler}},
  \bibinfo {author} {\bibfnamefont {S.}~\bibnamefont {Inouye}}, \bibinfo
  {author} {\bibfnamefont {A.~P.}\ \bibnamefont {Chikkatur}}, \bibinfo {author}
  {\bibfnamefont {S.~R.}\ \bibnamefont {Leslie}}, \bibinfo {author}
  {\bibfnamefont {K.~L.}\ \bibnamefont {Moore}}, \bibinfo {author}
  {\bibfnamefont {V.}~\bibnamefont {Savalli}}, \ and\ \bibinfo {author}
  {\bibfnamefont {D.~M.}\ \bibnamefont {Stamper-Kurn}},\ }\href {\doibase
  10.1103/PhysRevLett.95.050401} {\bibfield  {journal} {\bibinfo  {journal}
  {Phys. Rev. Lett.}\ }\textbf {\bibinfo {volume} {95}},\ \bibinfo {pages}
  {050401} (\bibinfo {year} {2005})}\BibitemShut {NoStop}%
\bibitem [{\citenamefont {Bloch}\ \emph {et~al.}(2000)\citenamefont {Bloch},
  \citenamefont {Hansch},\ and\ \citenamefont {Esslinger}}]{Bloch2000a}%
  \BibitemOpen
  \bibfield  {author} {\bibinfo {author} {\bibfnamefont {I.}~\bibnamefont
  {Bloch}}, \bibinfo {author} {\bibfnamefont {T.~W.}\ \bibnamefont {Hansch}}, \
  and\ \bibinfo {author} {\bibfnamefont {T.}~\bibnamefont {Esslinger}},\ }\href
  {http://dx.doi.org/10.1038/35003132} {\bibfield  {journal} {\bibinfo
  {journal} {Nature}\ }\textbf {\bibinfo {volume} {403}},\ \bibinfo {pages}
  {166} (\bibinfo {year} {2000})}\BibitemShut {NoStop}%
\bibitem [{\citenamefont {Donner}\ \emph {et~al.}(2007)\citenamefont {Donner},
  \citenamefont {Ritter}, \citenamefont {Bourdel}, \citenamefont {{\"O}ttl},
  \citenamefont {K{\"o}hl},\ and\ \citenamefont {Esslinger}}]{Donner2007a}%
  \BibitemOpen
  \bibfield  {author} {\bibinfo {author} {\bibfnamefont {T.}~\bibnamefont
  {Donner}}, \bibinfo {author} {\bibfnamefont {S.}~\bibnamefont {Ritter}},
  \bibinfo {author} {\bibfnamefont {T.}~\bibnamefont {Bourdel}}, \bibinfo
  {author} {\bibfnamefont {A.}~\bibnamefont {{\"O}ttl}}, \bibinfo {author}
  {\bibfnamefont {M.}~\bibnamefont {K{\"o}hl}}, \ and\ \bibinfo {author}
  {\bibfnamefont {T.}~\bibnamefont {Esslinger}},\ }\href {\doibase
  10.1126/science.1138807} {\bibfield  {journal} {\bibinfo  {journal}
  {Science}\ }\textbf {\bibinfo {volume} {315}},\ \bibinfo {pages} {1556}
  (\bibinfo {year} {2007})}\BibitemShut {NoStop}%
\bibitem [{\citenamefont {Clad\'e}\ \emph {et~al.}(2009)\citenamefont
  {Clad\'e}, \citenamefont {Ryu}, \citenamefont {Ramanathan}, \citenamefont
  {Helmerson},\ and\ \citenamefont {Phillips}}]{Clade2009a}%
  \BibitemOpen
  \bibfield  {author} {\bibinfo {author} {\bibfnamefont {P.}~\bibnamefont
  {Clad\'e}}, \bibinfo {author} {\bibfnamefont {C.}~\bibnamefont {Ryu}},
  \bibinfo {author} {\bibfnamefont {A.}~\bibnamefont {Ramanathan}}, \bibinfo
  {author} {\bibfnamefont {K.}~\bibnamefont {Helmerson}}, \ and\ \bibinfo
  {author} {\bibfnamefont {W.~D.}\ \bibnamefont {Phillips}},\ }\href {\doibase
  10.1103/PhysRevLett.102.170401} {\bibfield  {journal} {\bibinfo  {journal}
  {Phys. Rev. Lett.}\ }\textbf {\bibinfo {volume} {102}},\ \bibinfo {pages}
  {170401} (\bibinfo {year} {2009})}\BibitemShut {NoStop}%
\bibitem [{\citenamefont {Navon}\ \emph {et~al.}(2015)\citenamefont {Navon},
  \citenamefont {Gaunt}, \citenamefont {Smith},\ and\ \citenamefont
  {Hadzibabic}}]{Navon2015a}%
  \BibitemOpen
  \bibfield  {author} {\bibinfo {author} {\bibfnamefont {N.}~\bibnamefont
  {Navon}}, \bibinfo {author} {\bibfnamefont {A.~L.}\ \bibnamefont {Gaunt}},
  \bibinfo {author} {\bibfnamefont {R.~P.}\ \bibnamefont {Smith}}, \ and\
  \bibinfo {author} {\bibfnamefont {Z.}~\bibnamefont {Hadzibabic}},\ }\href
  {\doibase 10.1126/science.1258676} {\bibfield  {journal} {\bibinfo  {journal}
  {Science}\ }\textbf {\bibinfo {volume} {347}},\ \bibinfo {pages} {167}
  (\bibinfo {year} {2015})}\BibitemShut {NoStop}%
\bibitem [{\citenamefont {Chomaz}\ \emph {et~al.}(2015)\citenamefont {Chomaz},
  \citenamefont {Corman}, \citenamefont {Bienaim{\'e}}, \citenamefont
  {Desbuquois}, \citenamefont {Weitenberg}, \citenamefont {Nascimb{\`e}ne},
  \citenamefont {Beugnon},\ and\ \citenamefont {Dalibard}}]{Chomaz2015a}%
  \BibitemOpen
  \bibfield  {author} {\bibinfo {author} {\bibfnamefont {L.}~\bibnamefont
  {Chomaz}}, \bibinfo {author} {\bibfnamefont {L.}~\bibnamefont {Corman}},
  \bibinfo {author} {\bibfnamefont {T.}~\bibnamefont {Bienaim{\'e}}}, \bibinfo
  {author} {\bibfnamefont {R.}~\bibnamefont {Desbuquois}}, \bibinfo {author}
  {\bibfnamefont {C.}~\bibnamefont {Weitenberg}}, \bibinfo {author}
  {\bibfnamefont {S.}~\bibnamefont {Nascimb{\`e}ne}}, \bibinfo {author}
  {\bibfnamefont {J.}~\bibnamefont {Beugnon}}, \ and\ \bibinfo {author}
  {\bibfnamefont {J.}~\bibnamefont {Dalibard}},\ }\href
  {http://dx.doi.org/10.1038/ncomms7162} {\bibfield  {journal} {\bibinfo
  {journal} {Nat Commun}\ }\textbf {\bibinfo {volume} {6}} (\bibinfo {year}
  {2015})}\BibitemShut {NoStop}%
\bibitem [{\citenamefont {Natu}\ and\ \citenamefont
  {Mueller}(2010)}]{Natu2010a}%
  \BibitemOpen
  \bibfield  {author} {\bibinfo {author} {\bibfnamefont {S.~S.}\ \bibnamefont
  {Natu}}\ and\ \bibinfo {author} {\bibfnamefont {E.~J.}\ \bibnamefont
  {Mueller}},\ }\href {\doibase 10.1103/PhysRevA.81.053617} {\bibfield
  {journal} {\bibinfo  {journal} {Phys. Rev. A}\ }\textbf {\bibinfo {volume}
  {81}},\ \bibinfo {pages} {053617} (\bibinfo {year} {2010})}\BibitemShut
  {NoStop}%
\bibitem [{\citenamefont {Bray}(1994)}]{Bray1994}%
  \BibitemOpen
  \bibfield  {author} {\bibinfo {author} {\bibfnamefont {A.}~\bibnamefont
  {Bray}},\ }\href {\doibase 10.1080/00018739400101505} {\bibfield  {journal}
  {\bibinfo  {journal} {Advances in Physics}\ }\textbf {\bibinfo {volume}
  {43}},\ \bibinfo {pages} {357} (\bibinfo {year} {1994})}\BibitemShut
  {NoStop}%
\bibitem [{\citenamefont {Stauffer}\ and\ \citenamefont
  {Aharony}(1994)}]{Stauffer1994}%
  \BibitemOpen
  \bibfield  {author} {\bibinfo {author} {\bibfnamefont {D.}~\bibnamefont
  {Stauffer}}\ and\ \bibinfo {author} {\bibfnamefont {A.}~\bibnamefont
  {Aharony}},\ }\href@noop {} {\emph {\bibinfo {title} {Introduction to
  Percolation Theory}}}\ (\bibinfo  {publisher} {Taylor \& Francis},\ \bibinfo
  {year} {1994})\BibitemShut {NoStop}%
\bibitem [{\citenamefont {Takeuchi}\ \emph {et~al.}(2015)\citenamefont
  {Takeuchi}, \citenamefont {Mizuno},\ and\ \citenamefont
  {Dehara}}]{Takeuchi2015a}%
  \BibitemOpen
  \bibfield  {author} {\bibinfo {author} {\bibfnamefont {H.}~\bibnamefont
  {Takeuchi}}, \bibinfo {author} {\bibfnamefont {Y.}~\bibnamefont {Mizuno}}, \
  and\ \bibinfo {author} {\bibfnamefont {K.}~\bibnamefont {Dehara}},\ }\href
  {\doibase 10.1103/PhysRevA.92.043608} {\bibfield  {journal} {\bibinfo
  {journal} {Phys. Rev. A}\ }\textbf {\bibinfo {volume} {92}},\ \bibinfo
  {pages} {043608} (\bibinfo {year} {2015})}\BibitemShut {NoStop}%
\bibitem [{\citenamefont {Takeuchi}(2016)}]{Takeuchi2016a}%
  \BibitemOpen
  \bibfield  {author} {\bibinfo {author} {\bibfnamefont {H.}~\bibnamefont
  {Takeuchi}},\ }\href {\doibase 10.1007/s10909-016-1543-7} {\bibfield
  {journal} {\bibinfo  {journal} {Journal of Low Temperature Physics}\ }\textbf
  {\bibinfo {volume} {183}},\ \bibinfo {pages} {169} (\bibinfo {year}
  {2016})}\BibitemShut {NoStop}%
\bibitem [{\citenamefont {Ho}(1998)}]{Ho1998a}%
  \BibitemOpen
  \bibfield  {author} {\bibinfo {author} {\bibfnamefont {T.-L.}\ \bibnamefont
  {Ho}},\ }\href {\doibase 10.1103/PhysRevLett.81.742} {\bibfield  {journal}
  {\bibinfo  {journal} {Phys. Rev. Lett.}\ }\textbf {\bibinfo {volume} {81}},\
  \bibinfo {pages} {742} (\bibinfo {year} {1998})}\BibitemShut {NoStop}%
\bibitem [{\citenamefont {Ohmi}\ and\ \citenamefont
  {Machida}(1998)}]{Ohmi1998a}%
  \BibitemOpen
  \bibfield  {author} {\bibinfo {author} {\bibfnamefont {T.}~\bibnamefont
  {Ohmi}}\ and\ \bibinfo {author} {\bibfnamefont {K.}~\bibnamefont {Machida}},\
  }\href {\doibase 10.1143/JPSJ.67.1822} {\bibfield  {journal} {\bibinfo
  {journal} {J. Phys. Soc. Jpn}\ }\textbf {\bibinfo {volume} {67}},\ \bibinfo
  {pages} {1822} (\bibinfo {year} {1998})}\BibitemShut {NoStop}%
\bibitem [{\citenamefont {Chang}\ \emph {et~al.}(2004)\citenamefont {Chang},
  \citenamefont {Hamley}, \citenamefont {Barrett}, \citenamefont {Sauer},
  \citenamefont {Fortier}, \citenamefont {Zhang}, \citenamefont {You},\ and\
  \citenamefont {Chapman}}]{Chang2004a}%
  \BibitemOpen
  \bibfield  {author} {\bibinfo {author} {\bibfnamefont {M.-S.}\ \bibnamefont
  {Chang}}, \bibinfo {author} {\bibfnamefont {C.~D.}\ \bibnamefont {Hamley}},
  \bibinfo {author} {\bibfnamefont {M.~D.}\ \bibnamefont {Barrett}}, \bibinfo
  {author} {\bibfnamefont {J.~A.}\ \bibnamefont {Sauer}}, \bibinfo {author}
  {\bibfnamefont {K.~M.}\ \bibnamefont {Fortier}}, \bibinfo {author}
  {\bibfnamefont {W.}~\bibnamefont {Zhang}}, \bibinfo {author} {\bibfnamefont
  {L.}~\bibnamefont {You}}, \ and\ \bibinfo {author} {\bibfnamefont {M.~S.}\
  \bibnamefont {Chapman}},\ }\href {\doibase 10.1103/PhysRevLett.92.140403}
  {\bibfield  {journal} {\bibinfo  {journal} {Phys. Rev. Lett.}\ }\textbf
  {\bibinfo {volume} {92}},\ \bibinfo {pages} {140403} (\bibinfo {year}
  {2004})}\BibitemShut {NoStop}%
\bibitem [{\citenamefont {Damle}\ \emph
  {et~al.}(1996{\natexlab{a}})\citenamefont {Damle}, \citenamefont {Senthil},
  \citenamefont {Majumdar},\ and\ \citenamefont {Sachdev}}]{damle1996b}%
  \BibitemOpen
  \bibfield  {author} {\bibinfo {author} {\bibfnamefont {K.}~\bibnamefont
  {Damle}}, \bibinfo {author} {\bibfnamefont {T.}~\bibnamefont {Senthil}},
  \bibinfo {author} {\bibfnamefont {S.~N.}\ \bibnamefont {Majumdar}}, \ and\
  \bibinfo {author} {\bibfnamefont {S.}~\bibnamefont {Sachdev}},\ }\href
  {http://stacks.iop.org/0295-5075/36/i=1/a=007} {\bibfield  {journal}
  {\bibinfo  {journal} {EPL (Europhysics Letters)}\ }\textbf {\bibinfo {volume}
  {36}},\ \bibinfo {pages} {7} (\bibinfo {year}
  {1996}{\natexlab{a}})}\BibitemShut {NoStop}%
\bibitem [{\citenamefont {Barnett}\ \emph {et~al.}(2011)\citenamefont
  {Barnett}, \citenamefont {Polkovnikov},\ and\ \citenamefont
  {Vengalattore}}]{Barnett2011}%
  \BibitemOpen
  \bibfield  {author} {\bibinfo {author} {\bibfnamefont {R.}~\bibnamefont
  {Barnett}}, \bibinfo {author} {\bibfnamefont {A.}~\bibnamefont
  {Polkovnikov}}, \ and\ \bibinfo {author} {\bibfnamefont {M.}~\bibnamefont
  {Vengalattore}},\ }\href {\doibase 10.1103/PhysRevA.84.023606} {\bibfield
  {journal} {\bibinfo  {journal} {Phys. Rev. A}\ }\textbf {\bibinfo {volume}
  {84}},\ \bibinfo {pages} {023606} (\bibinfo {year} {2011})}\BibitemShut
  {NoStop}%
\bibitem [{\citenamefont {Seo}\ \emph {et~al.}(2015)\citenamefont {Seo},
  \citenamefont {Kang}, \citenamefont {Kwon},\ and\ \citenamefont
  {Shin}}]{Seo2015a}%
  \BibitemOpen
  \bibfield  {author} {\bibinfo {author} {\bibfnamefont {S.~W.}\ \bibnamefont
  {Seo}}, \bibinfo {author} {\bibfnamefont {S.}~\bibnamefont {Kang}}, \bibinfo
  {author} {\bibfnamefont {W.~J.}\ \bibnamefont {Kwon}}, \ and\ \bibinfo
  {author} {\bibfnamefont {Y.-i.}\ \bibnamefont {Shin}},\ }\href {\doibase
  10.1103/PhysRevLett.115.015301} {\bibfield  {journal} {\bibinfo  {journal}
  {Phys. Rev. Lett.}\ }\textbf {\bibinfo {volume} {115}},\ \bibinfo {pages}
  {015301} (\bibinfo {year} {2015})}\BibitemShut {NoStop}%
\bibitem [{\citenamefont {Symes}\ \emph {et~al.}(2014)\citenamefont {Symes},
  \citenamefont {Baillie},\ and\ \citenamefont {Blakie}}]{Symes2014a}%
  \BibitemOpen
  \bibfield  {author} {\bibinfo {author} {\bibfnamefont {L.~M.}\ \bibnamefont
  {Symes}}, \bibinfo {author} {\bibfnamefont {D.}~\bibnamefont {Baillie}}, \
  and\ \bibinfo {author} {\bibfnamefont {P.~B.}\ \bibnamefont {Blakie}},\
  }\href {\doibase 10.1103/PhysRevA.90.053616} {\bibfield  {journal} {\bibinfo
  {journal} {Phys. Rev. A}\ }\textbf {\bibinfo {volume} {90}},\ \bibinfo
  {pages} {053616} (\bibinfo {year} {2014})}\BibitemShut {NoStop}%
\bibitem [{\citenamefont {Blakie}\ \emph {et~al.}(2008)\citenamefont {Blakie},
  \citenamefont {Bradley}, \citenamefont {Davis}, \citenamefont {Ballagh},\
  and\ \citenamefont {Gardiner}}]{cfieldRev2008}%
  \BibitemOpen
  \bibfield  {author} {\bibinfo {author} {\bibfnamefont {P.~B.}\ \bibnamefont
  {Blakie}}, \bibinfo {author} {\bibfnamefont {A.~S.}\ \bibnamefont {Bradley}},
  \bibinfo {author} {\bibfnamefont {M.~J.}\ \bibnamefont {Davis}}, \bibinfo
  {author} {\bibfnamefont {R.~J.}\ \bibnamefont {Ballagh}}, \ and\ \bibinfo
  {author} {\bibfnamefont {C.~W.}\ \bibnamefont {Gardiner}},\ }\href@noop {}
  {\bibfield  {journal} {\bibinfo  {journal} {Adv. Phys.}\ }\textbf {\bibinfo
  {volume} {57}},\ \bibinfo {pages} {363} (\bibinfo {year} {2008})}\BibitemShut
  {NoStop}%
\bibitem [{\citenamefont {Symes}\ \emph {et~al.}(2016)\citenamefont {Symes},
  \citenamefont {McLachlan},\ and\ \citenamefont {Blakie}}]{Symes2016a}%
  \BibitemOpen
  \bibfield  {author} {\bibinfo {author} {\bibfnamefont {L.~M.}\ \bibnamefont
  {Symes}}, \bibinfo {author} {\bibfnamefont {R.~I.}\ \bibnamefont
  {McLachlan}}, \ and\ \bibinfo {author} {\bibfnamefont {P.~B.}\ \bibnamefont
  {Blakie}},\ }\href {\doibase 10.1103/PhysRevE.93.053309} {\bibfield
  {journal} {\bibinfo  {journal} {Phys. Rev. E}\ }\textbf {\bibinfo {volume}
  {93}},\ \bibinfo {pages} {053309} (\bibinfo {year} {2016})}\BibitemShut
  {NoStop}%
\bibitem [{\citenamefont {Ao}\ and\ \citenamefont {Chui}(1998)}]{Ao1998a}%
  \BibitemOpen
  \bibfield  {author} {\bibinfo {author} {\bibfnamefont {P.}~\bibnamefont
  {Ao}}\ and\ \bibinfo {author} {\bibfnamefont {S.~T.}\ \bibnamefont {Chui}},\
  }\href {\doibase 10.1103/PhysRevA.58.4836} {\bibfield  {journal} {\bibinfo
  {journal} {Phys. Rev. A}\ }\textbf {\bibinfo {volume} {58}},\ \bibinfo
  {pages} {4836} (\bibinfo {year} {1998})}\BibitemShut {NoStop}%
\bibitem [{\citenamefont {Yurke}\ \emph {et~al.}(1993)\citenamefont {Yurke},
  \citenamefont {Pargellis}, \citenamefont {Kovacs},\ and\ \citenamefont
  {Huse}}]{Yurke1993a}%
  \BibitemOpen
  \bibfield  {author} {\bibinfo {author} {\bibfnamefont {B.}~\bibnamefont
  {Yurke}}, \bibinfo {author} {\bibfnamefont {A.~N.}\ \bibnamefont
  {Pargellis}}, \bibinfo {author} {\bibfnamefont {T.}~\bibnamefont {Kovacs}}, \
  and\ \bibinfo {author} {\bibfnamefont {D.~A.}\ \bibnamefont {Huse}},\ }\href
  {\doibase 10.1103/PhysRevE.47.1525} {\bibfield  {journal} {\bibinfo
  {journal} {Phys. Rev. E}\ }\textbf {\bibinfo {volume} {47}},\ \bibinfo
  {pages} {1525} (\bibinfo {year} {1993})}\BibitemShut {NoStop}%
\bibitem [{\citenamefont {Bray}\ \emph {et~al.}(2000)\citenamefont {Bray},
  \citenamefont {Briant},\ and\ \citenamefont {Jervis}}]{Bray2000a}%
  \BibitemOpen
  \bibfield  {author} {\bibinfo {author} {\bibfnamefont {A.~J.}\ \bibnamefont
  {Bray}}, \bibinfo {author} {\bibfnamefont {A.~J.}\ \bibnamefont {Briant}}, \
  and\ \bibinfo {author} {\bibfnamefont {D.~K.}\ \bibnamefont {Jervis}},\
  }\href {\doibase 10.1103/PhysRevLett.84.1503} {\bibfield  {journal} {\bibinfo
   {journal} {Phys. Rev. Lett.}\ }\textbf {\bibinfo {volume} {84}},\ \bibinfo
  {pages} {1503} (\bibinfo {year} {2000})}\BibitemShut {NoStop}%
\bibitem [{\citenamefont {Damle}\ \emph
  {et~al.}(1996{\natexlab{b}})\citenamefont {Damle}, \citenamefont {Majumdar},\
  and\ \citenamefont {Sachdev}}]{Damle1996a}%
  \BibitemOpen
  \bibfield  {author} {\bibinfo {author} {\bibfnamefont {K.}~\bibnamefont
  {Damle}}, \bibinfo {author} {\bibfnamefont {S.~N.}\ \bibnamefont {Majumdar}},
  \ and\ \bibinfo {author} {\bibfnamefont {S.}~\bibnamefont {Sachdev}},\ }\href
  {\doibase 10.1103/PhysRevA.54.5037} {\bibfield  {journal} {\bibinfo
  {journal} {Phys. Rev. A}\ }\textbf {\bibinfo {volume} {54}},\ \bibinfo
  {pages} {5037} (\bibinfo {year} {1996}{\natexlab{b}})}\BibitemShut {NoStop}%
\bibitem [{\citenamefont {{Karl}}\ and\ \citenamefont
  {{Gasenzer}}(2016)}]{Karl2016a}%
  \BibitemOpen
  \bibfield  {author} {\bibinfo {author} {\bibfnamefont {M.}~\bibnamefont
  {{Karl}}}\ and\ \bibinfo {author} {\bibfnamefont {T.}~\bibnamefont
  {{Gasenzer}}},\ }\href@noop {} {\bibfield  {journal} {\bibinfo  {journal}
  {ArXiv e-prints}\ } (\bibinfo {year} {2016})},\ \Eprint
  {http://arxiv.org/abs/1611.01163} {arXiv:1611.01163 [cond-mat.quant-gas]}
  \BibitemShut {NoStop}%
\bibitem [{\citenamefont {Hoshen}\ and\ \citenamefont
  {Kopelman}(1976)}]{Hoshen1976a}%
  \BibitemOpen
  \bibfield  {author} {\bibinfo {author} {\bibfnamefont {J.}~\bibnamefont
  {Hoshen}}\ and\ \bibinfo {author} {\bibfnamefont {R.}~\bibnamefont
  {Kopelman}},\ }\href {\doibase 10.1103/PhysRevB.14.3438} {\bibfield
  {journal} {\bibinfo  {journal} {Phys. Rev. B}\ }\textbf {\bibinfo {volume}
  {14}},\ \bibinfo {pages} {3438} (\bibinfo {year} {1976})}\BibitemShut
  {NoStop}%
\bibitem [{\citenamefont {Furukawa}(1985)}]{Furukawa1985}%
  \BibitemOpen
  \bibfield  {author} {\bibinfo {author} {\bibfnamefont {H.}~\bibnamefont
  {Furukawa}},\ }\href {\doibase 10.1103/PhysRevA.31.1103} {\bibfield
  {journal} {\bibinfo  {journal} {Phys. Rev. A}\ }\textbf {\bibinfo {volume}
  {31}},\ \bibinfo {pages} {1103} (\bibinfo {year} {1985})}\BibitemShut
  {NoStop}%
\bibitem [{\citenamefont {Fujimoto}\ and\ \citenamefont
  {Tsubota}(2016)}]{Fujimoto2016a}%
  \BibitemOpen
  \bibfield  {author} {\bibinfo {author} {\bibfnamefont {K.}~\bibnamefont
  {Fujimoto}}\ and\ \bibinfo {author} {\bibfnamefont {M.}~\bibnamefont
  {Tsubota}},\ }\href {\doibase 10.1103/PhysRevA.93.033620} {\bibfield
  {journal} {\bibinfo  {journal} {Phys. Rev. A}\ }\textbf {\bibinfo {volume}
  {93}},\ \bibinfo {pages} {033620} (\bibinfo {year} {2016})}\BibitemShut
  {NoStop}%
\end{thebibliography}

%

%
\end{document}